\documentclass[]{jfm}

\usepackage{graphicx}
\usepackage{newtxtext}
\usepackage{newtxmath}
\usepackage{natbib}
\usepackage{amsmath}
\usepackage{amssymb}
\usepackage{graphicx}
\usepackage{epsfig}
\usepackage{hyperref}
\usepackage{bm}
\usepackage{comment}
\usepackage{subcaption}
\usepackage{float}
\hypersetup{
    colorlinks = true,
    urlcolor   = blue,
    citecolor  = black,
}

\newcommand{\RomanNumeralCaps}[1]

\graphicspath{{figures/}}
\newcommand{\ddd}{\mathrm{d}}

\title{Mode-to-mode nonlinear energy transfer in turbulent channel flows}

\author{Jitong Ding \aff{1}
  \corresp{\email{jitongd@student.unimelb.edu.au}},
  Daniel Chung \aff{1}
 \and Simon J. Illingworth \aff{1}}

\affiliation{\aff{1}Department of Mechanical Engineering, University of Melbourne, Victoria 3010, Australia
}

\begin{document}
\maketitle

\begin{abstract}
We investigate nonlinear energy transfer for channel flows at friction Reynolds numbers of $Re_{\tau}=180$ and $590$.
The key feature of the analysis is that we introduce a new variable, which quantifies the energy transferred from a source mode to a recipient mode through explicit examination of nonlinear triadic interactions in streamwise-spanwise wavenumber space.
First, we use this variable to quantify the nonlinear energy transfer gain and loss for individual Fourier modes.
The nonlinear energy transfer gain and loss cannot be directly obtained from the turbulent kinetic energy (TKE) equation.
Second, we quantify the nonlinear energy transfer budgets for three types of structures: streamwise streaks, oblique waves and Tollmien-Schlichting waves. 
We found that a transverse cascade from streamwise-elongated modes to spanwise-elongated modes exists in all three structures.
Third, we quantify the forward and inverse cascades between resolved scales and subgrid scales in the spirit of large-eddy simulation. 
For the cutoff wavelength range we consider, the forward and inverse cascades between the resolved scales and subgrid scales result in a net forward cascade from the resolved scales to the subgrid scales. 
The shape of the net forward cascade curve with respect to the cutoff wavelength resembles the net forward cascade predicted by the Smagorinsky eddy viscosity.
\end{abstract}



\section{Introduction}
\label{sec:headings}
Turbulence involves energy transfer across a wide range of scales and is important for many applications, such as combustion \citep{ertesvaag2000eddy}, ocean engineering \citep{hasselmann1963bispectra} and the earth's climate \citep{richardson1922weather}. 
It is generally understood that energy is transferred from large scales to small scales, known as the forward cascade \citep{richardson1922weather, kolmogorov1941local}. 
But an inverse cascade in which energy is transferred from small scales to large scales also occurs \citep{domaradzki1994energy, dunn2003anisotropy, cimarelli2013paths, cimarelli2016cascades}.
Both forward and inverse cascades are involved in vortex regeneration in the self-sustaining process for wall-turbulence \citep{hamilton1995regeneration, waleffe1997self}. 
In turbulence modelling and simulation, failure to consider backscatter can result in inaccurate large-eddy simulations \citep{piomelli1991subgrid, hartel1994subgrid, cimarelli2014physics}.
The above considerations are all dictated by the physics of the energy cascade across scales in wall-bounded turbulent flows.

This energy cascade can be viewed in either physical space or Fourier space. 
In physical space, the energy cascade is characterised by high spatio-temporal intermittency \citep{meneveau1991multifractal}. 
In Fourier space, for flow geometries (such as channels, pipes, boundary layers) with at least one homogeneous spatial dimension, the energy cascade can be interpreted as energy redistributed among different Fourier modes, described by the nonlinear energy transfer term in the spectral turbulent kinetic energy (sTKE) equation \citep{tennekes1972first, pope2000turbulent}. 
Here, a mode refers to scales with a particular wavelength in the homogeneous direction. 
The nonlinear energy transfer term in the spectral energy budget represents the net energy received by a particular mode from all other modes through nonlinear interactions since the term is a convolution of all wavenumber-compatible triadic interactions. 
Triadic interaction refers to the energy transfer among three modes whose wavenumbers satisfy $\bm{k}+\bm{p}+\bm{q}=\bm{0}$ \citep{domaradzki1990local, domaradzki1992nonlocal, waleffe1992nature, domaradzki1994energy}.
However, we merely know the net energy transfer for each Fourier mode but not the detailed contributions to this net value from the nonlinear energy transfer term, because the convolution hides the individual triadic interactions.

Compared to homogeneous isotropic turbulence, only a few studies have explored the triadic interactions of nonlinear energy transfer for channel flows \citep{domaradzki1994energy, webber2002energy, cho2018scale, karban2023modal}. 
A single wavenumber triad could be interpreted as energy transfer from a source mode to a recipient mode via an advective mode \citep{domaradzki1990local, smyth1992spectral, webber2002energy, alexakis2005imprint, jin2021energy}.
To gain more insight into the energy cascade in channel flows, we expand the convolution of the nonlinear energy transfer term in streamwise-spanwise wavenumber space.
Following this, we formulate a new variable $\hat{M}_{(s_x,s_y)(k_x,k_y)}$ (defined in \S\ref{section_Nonlinear_energy_transfer_between_two_modes}) that represents mode-to-mode nonlinear energy transfer in streamwise-spanwise wavenumber space. 
Compared to the convolution term which only gives us the net energy transfer for a specific Fourier mode, with this variable, we are able to quantify the energy transfer between any two Fourier modes. 

We use this four-dimensional variable $\hat{M}_{(s_x,s_y)(k_x,k_y)}$ (one source mode and one recipient mode with each mode consisting of a streamwise wavenumber and a spanwise wavenumber) to explore three things using direct numerical simulation (DNS) datasets at $Re_{\tau}=180$ and $590$.
First, since the nonlinear energy transfer term in the sTKE equations could only give us one value representing the net energy transfer for each mode previously, we use this new variable $\hat{M}_{(s_x,s_y)(k_x,k_y)}$ to obtain two additional values quantifying the net energy transfer gain and loss due to nonlinear interactions for each mode.
Second, we investigate the nonlinear energy transfer budgets for three types of structures: streamwise streaks, oblique waves and Tollmien–Schlichting (TS) waves which are important in sustaining turbulence in wall-bounded flows.
Third, similar to a large-eddy simulation set-up, we quantify the forward cascade and inverse cascade between resolved and subgrid scales. 
We further compare the forward cascade calculated using $\hat{M}$ with the forward cascade calculated using the Smagorinsky eddy viscosity.
The similarities and differences between $Re_{\tau}=180$ and $590$ are discussed. 

This paper is organised as follows. 
In \S\ref{section_methods}, the equations for spectral turbulent kinetic energy (sTKE) budget and the new variable $\hat{M}_{(s_x,s_y)(k_x,k_y)}$ representing mode-to-mode nonlinear energy transfer are derived. 
The DNS datasets are described in \S\ref{section_flow_descriptions}. 
\S\ref{section_results} presents the results.
Specifically, \S\ref{section_PDN} revisits the previous study from \citet{symon2021energy} about the wall-normal integrated spectral energy transfer budget; 
\S\ref{section_3_examples} uses two examples to interpret the new variable and to illustrate energy transfer pathways;
\S\ref{section_source_recipient} presents the positive and negative nonlinear energy transfer spectra;
\S\ref{section_three_structures} investigates the nonlinear energy transfer of streamwise streaks, oblique waves and TS waves;
\S\ref{section_forward_inverse_cascade} quantifies the forward cascade and inverse cascade between resolved scales and subgrid scales in the spirit of LES. 
Conclusions are drawn in \S\ref{section_conclusions}.

\section{Methods}\label{section_methods}
In \S \ref{section_plane_Poiseuille_flow_equations}, we describe the governing equations for plane Poiseuille flow and their non-dimensionalisation. Then, an introduction to the spectral turbulent kinetic energy equation integrated across the channel height is presented in \S \ref{section_Spectral_turbulent_kinetic_energy_equation}. In \S \ref{section_Nonlinear_energy_transfer_between_two_modes}, we introduce a four-dimensional variable quantifying mode-to-mode nonlinear energy transfer and the pertinent properties.

\subsection{Plane Poiseuille flow equations} \label{section_plane_Poiseuille_flow_equations}
Consider the non-dimensional incompressible Navier-Stokes equations for the fluctuation velocities after a Reynolds decomposition, $\mathcal{U}_i = U_i + u_i$, where $\mathcal{U}_i$, $U_i$ and $u_i$ represent the instantaneous velocity, time-averaged velocity and fluctuation velocity, respectively: 
\begin{equation}
\begin{split}
\frac{\partial u_i}{\partial x_i} &= 0  \\
\frac{\partial u_i}{\partial t} + u_j\frac{\partial U_i}{\partial x_j} + U_j\frac{\partial u_i}{\partial x_j} + \frac{\partial}{\partial x_j} (u_i u_j- \overline{u_i u_j}) &= -\frac{\partial p}{\partial x_i} + \frac{1}{Re_{\tau}} \frac{\partial^2 u_i}{\partial x_j \partial x_j}
\end{split}
\label{eqa_NSEs}
\end{equation}
where the indices $i=1,2,3$ represent the $x$ (streamwise), $y$ (spanwise) and $z$ (wall-normal) directions. 
The corresponding velocity components are denoted by $u$, $v$ and $w$. 
Pressure is denoted as $p$. 
Length scales are non-dimensionalised using the channel half-height $h$, time scales are non-dimensionalised using $h/u_{\tau}$ and pressure is non-dimensionalised using $\rho u_{\tau}^2$, where $u_{\tau}=\sqrt{\tau_w/\rho}$, $\rho$ is the density, $\tau_w$ is the mean wall shear stress and $u_{\tau}$ is the friction velocity. 
Then, the friction Reynolds number, $Re_{\tau} = hu_{\tau}/\nu$, is defined using $h$, $u_{\tau}$ and the kinematic viscosity, $\nu$. 
An overbar denotes the time-averaging operator.

\subsection{Spectral energy transfer budget} \label{section_Spectral_turbulent_kinetic_energy_equation}
Owing to the periodic assumption in the streamwise and spanwise directions, we investigate energy transfer in two-dimensional streamwise-spanwise wavenumber space. 
We define the inner product $\langle c_1,c_2 \rangle = \frac{1}{2} \int \limits_{-1}^{1} c_1^* c_2 \; \ddd z$, where $c_1$ and $c_2$ are two complex vectors, $*$ denotes the complex conjugate. 
We use this inner product definition to represent the wall-normal integrated kinetic energy at mode $(k_x,k_y)$ \citep{reddy1993energy, domaradzki1994energy}: 
\begin{equation}
\hat{E}(k_x,k_y) = 
\frac{1}{2} \langle \hat{u}^{(k_x,k_y)}, \hat{u}^{(k_x,k_y)} \rangle +
\frac{1}{2} \langle \hat{v}^{(k_x,k_y)}, \hat{v}^{(k_x,k_y)} \rangle + 
\frac{1}{2} \langle \hat{w}^{(k_x,k_y)}, \hat{w}^{(k_x,k_y)} \rangle 
\label{eqa_kinetic_energy_kxky}
\end{equation}
where $k_x$ is the streamwise wavenumber and $k_y$ is the spanwise wavenumber. 
The superscript refers to the individual Fourier mode under consideration: $\hat{u}^{(k_x,k_y)}$ is the Fourier coefficient of $u_i$ at wavenumber $(k_x,k_y)$.

The spectral energy transfer budget can be obtained by first taking Fourier transforms of equation \eqref{eqa_NSEs} in the $x$ and $y$ directions, and then multiplying by the conjugate mode $(\hat{u}_i^{(k_x,k_y)})^* = \hat{u}_i^{(-k_x,-k_y)}$. 
Then, we integrate the energy transfer budget in the wall-normal direction and obtain \citep{symon2021energy}:

\begin{subequations}
\begin{align}
&\overline{\frac{\partial \hat{E}{(k_x , k_y)}}{\partial t}} = \underbrace{-\overline{\langle \hat{u}, \frac{\ddd U}{\ddd z}\hat{w} \rangle} }_{\hat{P}{(k_x , k_y)}} 
-\underbrace{\frac{1}{Re_{\tau}} \overline{\langle \widehat{\frac{\partial u_i}{\partial x_j}},  \widehat{\frac{\partial u_i}{\partial x_j}} \rangle}}_{\hat{D} (k_x , k_y)}
\underbrace{-  \overline{\langle \hat{u}_i, \widehat{\frac{\partial u_i u_j}{\partial x_j}} \rangle}}_{\hat{N}(k_x , k_y)}  \label{eqa_PDN} \\
&\hat{P}{(k_x , k_y)} - \hat{D}{(k_x , k_y)} + \hat{N}{(k_x , k_y)} =0 \label{eqa_PDN0}
\end{align}
\end{subequations}
\noindent with summation implied in the coordinate directions over the repeating index $i$ or $j$. 
Equation \eqref{eqa_PDN} describes the wall-normal integrated energy transfer balance for a single Fourier mode. 
The left-hand side is the time derivative of the turbulent kinetic energy for a single Fourier mode. 
$\hat{P}$ represents production; $-\hat{D}$ represents (pseudo) dissipation \citep{pope2000turbulent}; and $\hat{N}$ represents the net nonlinear energy transfer (the net energy that mode $(k_x,k_y)$ receives through nonlinear interactions with all other modes). 
The size of an eddy corresponding to a given Fourier mode can be defined using the isotropic wall-parallel wavelength: $\lambda_I = 2\pi/k_I$, where $k_I^2 = k_x^2 + k_y^2$ \citep{jimenez2018coherent,lee2019spectral}.
However, this definition neglects the anisotropy in channel flows. 
For a statistically stationary flow, the left-hand side of equation \eqref{eqa_PDN} is zero, meaning that the wall-normal integrated production, dissipation and net nonlinear energy transfer reach a balance for each mode, as shown in equation \eqref{eqa_PDN0}. 
According to the normalisation described in \S\ref{section_plane_Poiseuille_flow_equations}, the energy transfer terms ($\hat{P}$, $\hat{D}$ and $\hat{N}$ in \eqref{eqa_PDN}) are non-dimensionalised by $u_{\tau}^3/h$.
Strictly speaking, production, dissipation and nonlinear energy transfer in equation \eqref{eqa_PDN} are energy transfer rates because it is the kinetic energy variation rate on the left-hand side of \eqref{eqa_PDN}.

Throughout this paper, we use non-negative wavenumbers to describe a mode $(k_x,k_y)$, where $k_x \geq 0$ and $k_y \geq 0$.
The energy transfer at mode $(k_x,k_y)$ with $k_x > 0$ and $k_y > 0$ contains the contributions from the wavenumber pairs $(k_x,k_y)$, $(-k_x,-k_y)$, $(-k_x,k_y)$ and $(k_x,-k_y)$.
For example, production at mode $(k_x,k_y)$ is equal to $\hat{P}(k_x,k_y)+\hat{P}(-k_x,k_y) + \mathrm{c.c.}$, where $\mathrm{c.c.}$ represents complex conjugate. 
The energy transfer at mode $(k_x,0)$ with $k_x > 0$ contains the contributions from the wavenumber pairs $(k_x,0)$ and $(-k_x,0)$.
For example, production at mode $(k_x,0)$ is equal to $\hat{P}(k_x,0) + \mathrm{c.c.}$. 
The energy transfer at mode $(0,k_y)$ with $k_y > 0$ contains the contributions from the wavenumber pairs $(0,k_y)$ and $(0,-k_y)$.
For example, production at mode $(0,k_y)$ is equal to $\hat{P}(0,k_y) + \mathrm{c.c.}$.
Following this, the energy transfer terms $\hat{P},\hat{D},\hat{N}$ for each mode $(k_x,k_y)$ with $k_x\geq0,k_y\geq0$ are real numbers.
The derivation of equation \eqref{eqa_PDN} can be found in \color{blue} \citet{ding2024nonlinear}.  \color{black}

$\hat{N}(k_x,k_y)$ in \eqref{eqa_PDN} is conservative:
\begin{equation}
\sum_{k_x} \sum_{k_y} \hat{N}(k_x,k_y) = 0 
    \label{eqa_N_0}
\end{equation}
Equation \eqref{eqa_N_0} states that the sum of the net nonlinear energy transfer across all Fourier modes is zero. 
This implies that nonlinear energy transfer redistributes energy across scales without adding or removing energy overall. 
This can also be seen from the Reynolds-Orr equation: the only energy source for turbulence is production and the only energy sink for turbulence is dissipation \citep{schmid2002stability}.

\subsection{Mode-to-mode nonlinear energy transfer $\hat{M}_{(s_x,s_y)(k_x,k_y)}$} \label{section_Nonlinear_energy_transfer_between_two_modes}
$\hat{N}(k_x,k_y)$ on the right-hand side of equation \eqref{eqa_PDN} represents the net energy received by mode $(k_x,k_y)$ from nonlinear interactions between mode  $(k_x,k_y)$ and all other modes, without giving information about the individual contributions to this net value. 
To illustrate, mode $(k_{x},k_{y})$ could potentially gain energy from mode $(s_{x1},s_{y1})$, lose energy to mode $(s_{x2},s_{y2})$, gain energy from mode $(s_{x3},s_{y3})$, lose energy to mode $(s_{x4},s_{y4})$, and so on. 
But those mode-to-mode nonlinear energy transfer is hidden because $\hat{N}$ represents a sum over all modes. 

$\hat{N}(k_x,k_y)$ in equation \eqref{eqa_PDN} can be expressed as a convolution composed of all wavenumber-compatible triadic interactions:
\begin{equation}
\hat{N}(k_x,k_y) =  \sum_{s_x} \sum_{s_y} \hat{M}_{(s_x,s_y)(k_x,k_y)} 
\label{eqa_N_convolution}
\end{equation}
where 
\begin{equation}
\hat{M}_{(s_x,s_y)(k_x,k_y)} = -  \overline{\hat{u}_i^{(-k_x,-k_y)}  \hat{u}_j^{(k_x - s_x,k_y- s_y)} \widehat{\frac{\partial u_i}{\partial x_j}}^{(s_x,s_y)}}
\label{eqa_M}
\end{equation}

\noindent A single set of triadic interaction involving three distinct modes can be understood as the energy transferred nonlinearly from mode $(s_x,s_y)$ to mode $(k_x,k_y)$ with the help of mode $(k_x - s_x,  k_y - s_y)$ \citep{domaradzki1990local, smyth1992spectral, webber2002energy, alexakis2005imprint, jin2021energy}. $s_x$ is the streamwise wavenumber and $s_y$ is the spanwise wavenumber for another mode $(s_x,s_y)$ different from mode $(k_x,k_y)$.
Thus, the four-dimensional variable $\hat{M}_{(s_x,s_y)(k_x,k_y)}$ describes the energy transferred nonlinearly from one Fourier mode $(s_x,s_y)$ to another Fourier mode $(k_x,k_y)$.
As mentioned before, we discuss modes composed of non-negative wavenumbers. 
The expressions for $\hat{M}_{(s_x,s_y)(k_x,k_y)}$ with $s_x,s_y,k_x,k_y \geq 0$ are discussed in detail in Appendix \ref{appendix_B}.
Most result discussions in this paper centre on the mode-to-mode nonlinear energy transfer $\hat{M}$ \eqref{eqa_M}. 
For the ease of reading, we sometimes omit the word `nonlinear'. 
Apart from production and dissipation, energy transfer, gaining energy and losing energy in the following discussions refer to $\hat{M}$ \eqref{eqa_M} due to nonlinear interactions.

Furthermore, $\hat{M}$ which represents wall-normal integrated energy transfer satisfies the following identities:
\begin{equation}
\hat{M}_{(s_x,s_y)(k_x,k_y)} = - \hat{M}_{(k_x,k_y)(s_x,s_y)}
\label{eqa_energy_transfer_mechanism_opposite}
\end{equation}
\begin{equation}
\hat{M}_{(k_x,k_y)(k_x,k_y)} = 0
\label{eqa_energy_transfer_mechanism_0}
\end{equation}
Equation \eqref{eqa_energy_transfer_mechanism_opposite} states that the nonlinear energy transfer from mode $(s_x,s_y)$ to mode $(k_x,k_y)$ is equal and opposite to the nonlinear energy transfer from mode $(k_x,k_y)$ to mode $(s_x,s_y)$. 
Equation \eqref{eqa_energy_transfer_mechanism_0} states that each mode transfers zero energy nonlinearly to itself. 
Equation \eqref{eqa_energy_transfer_mechanism_opposite} and equation \eqref{eqa_energy_transfer_mechanism_0} arise from the continuity equation and the boundary conditions at the walls (the proof can be found in \citet{ding2024nonlinear}).

We introduce the energy transfer quantities $\hat{P}, \hat{D}, \hat{N}, \hat{M}$ at discrete wavenumbers. 
The energy transfer could also be interpreted as energy transfer spectra. 
We use lowercase letters to represent the energy transfer spectra. 
For example, the production spectrum is $\hat{p}$ and is obtained through $\hat{p} = \frac{\hat{P}}{\Delta k_x \Delta k_y}$.
The energy transfer spectrum is usually visualised using the premultiplied form $k_x k_y \hat{p}$ on a logarithmic axis.

\section{Flow descriptions} \label{section_flow_descriptions}
Direct numerical simulations are performed using a staggered-grid fourth-order finite-difference solver \citep{chung2014idealised}. 
Table \ref{table_DNS} summarises the simulation parameters. 
For $L_x = 2\pi$ and $L_y=\pi$, the maximum wavenumbers resolved by the simulation are $k_x = \pm 55,\;k_y = \pm 110$ for $Re_{\tau} = 180$ and $k_x = \pm 191,\;k_y = \pm 382$ for $Re_{\tau} = 590$;
the minimum wavenumbers are $k_x = \pm 1,\;k_y = \pm 2$ for both Reynolds numbers.

\setlength{\tabcolsep}{8pt} 
\renewcommand{\arraystretch}{1.2} 

\begin{figure}
\centering
\begin{subfigure}[h]{.48\textwidth}
  \includegraphics[width=1.0\linewidth]{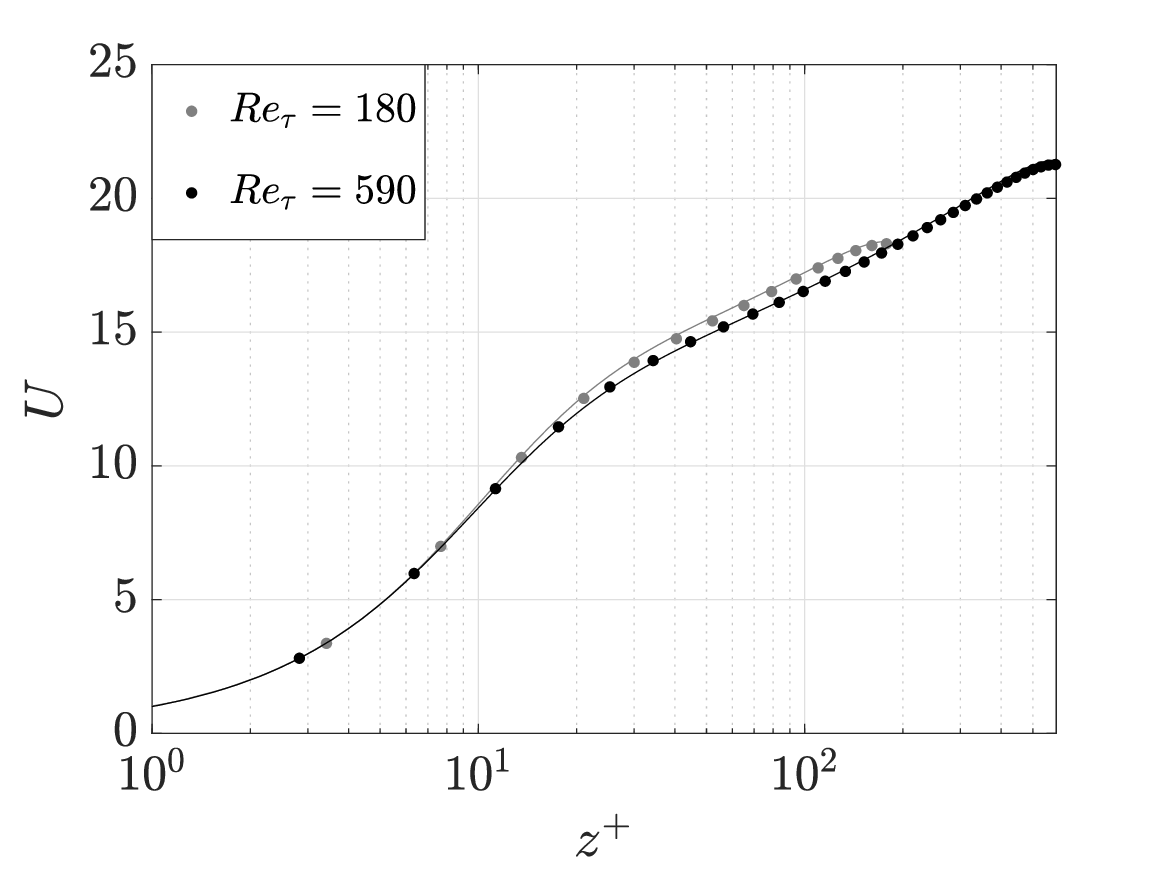}
\end{subfigure}
\begin{subfigure}[h]{.48\textwidth}
  \centering
  \includegraphics[width=1.0\linewidth]{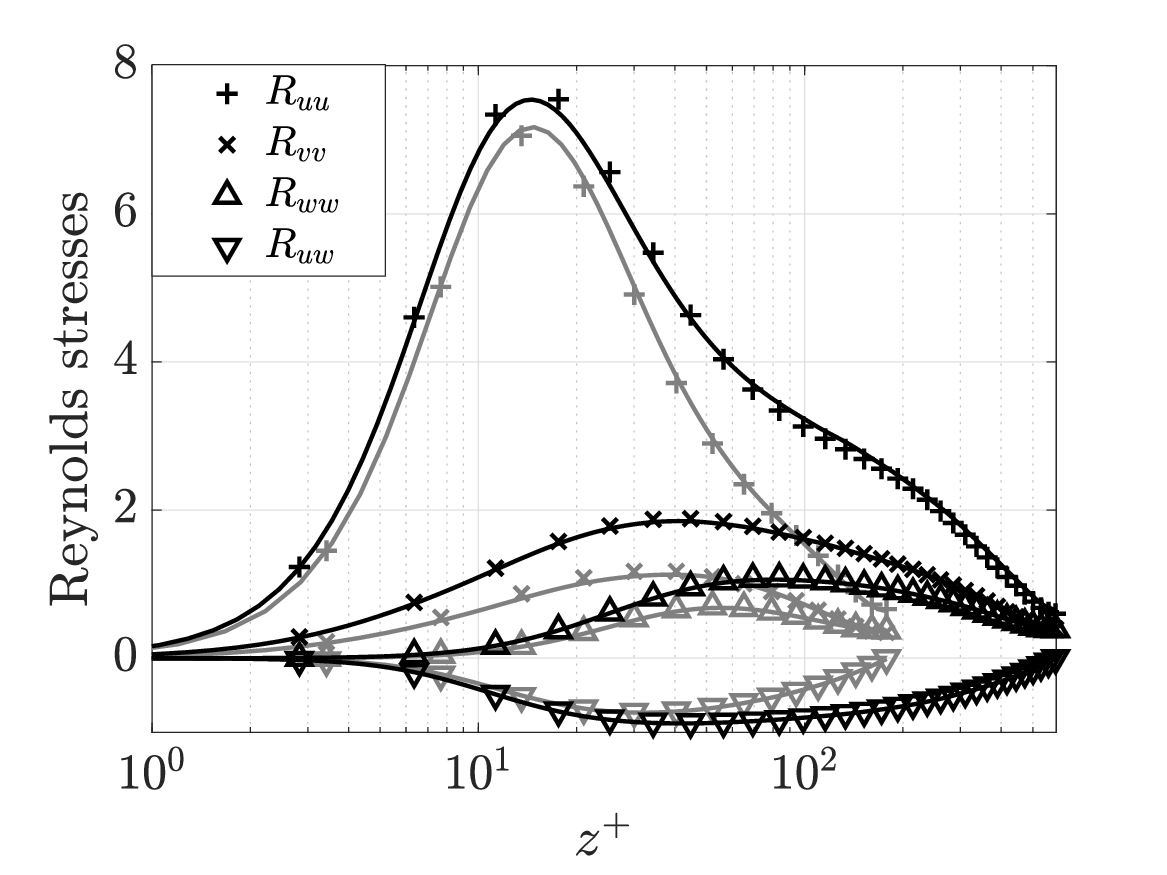}
\end{subfigure}
\caption{Comparison between the DNS dataset represented by solid lines and the standard DNS dataset \protect \citep{moser1999direct} represented by discrete markers. (a) Mean streamwise velocity. (b) Turbulence stresses. 
Grey, $Re_{\tau}=180$; black, $Re_{\tau}=590$. }
\label{fig_DNS_dataset}
\end{figure}

\begin{table}
\centering
\begin{tabular}{lllllllll}
\hline
$Re_{\tau}$ & $L_x$  & $L_y$ & $n_x \times n_y \times n_z$ & $\Delta x^+$ & $\Delta y^+$ & $\Delta z^+_{max}$ & $\Delta z^+_{min}$ & $\Delta t^+$ \\ 
180         & $2\pi$ & $\pi$ & $112 \times 112 \times 150$ & 10.00        & 5.05         & 3.77               & 0.04                  & 0.36      \\ 
590         & $2\pi$ & $\pi$ & $384 \times 384 \times 500$ & 9.65        & 4.83         & 3.71               & 0.01                  & 0.12       \\
\hline
\end{tabular}
\caption{Parameter setup of the DNS. $L$, domain length; $n$, number of grid points; $\Delta^+$, grid-spacing in viscous units; $\Delta t$ is the simulation time step.}
\label{table_DNS}
\end{table}
The time-averaged first-order and second-order statistics of the present DNS data show good agreement with the \citet{moser1999direct}, as shown in figure \ref{fig_DNS_dataset}.

\section{Results} \label{section_results}

\subsection{Energy transfer distributions} \label{section_PDN}
\begin{figure}
\centering
\includegraphics[scale=0.40]{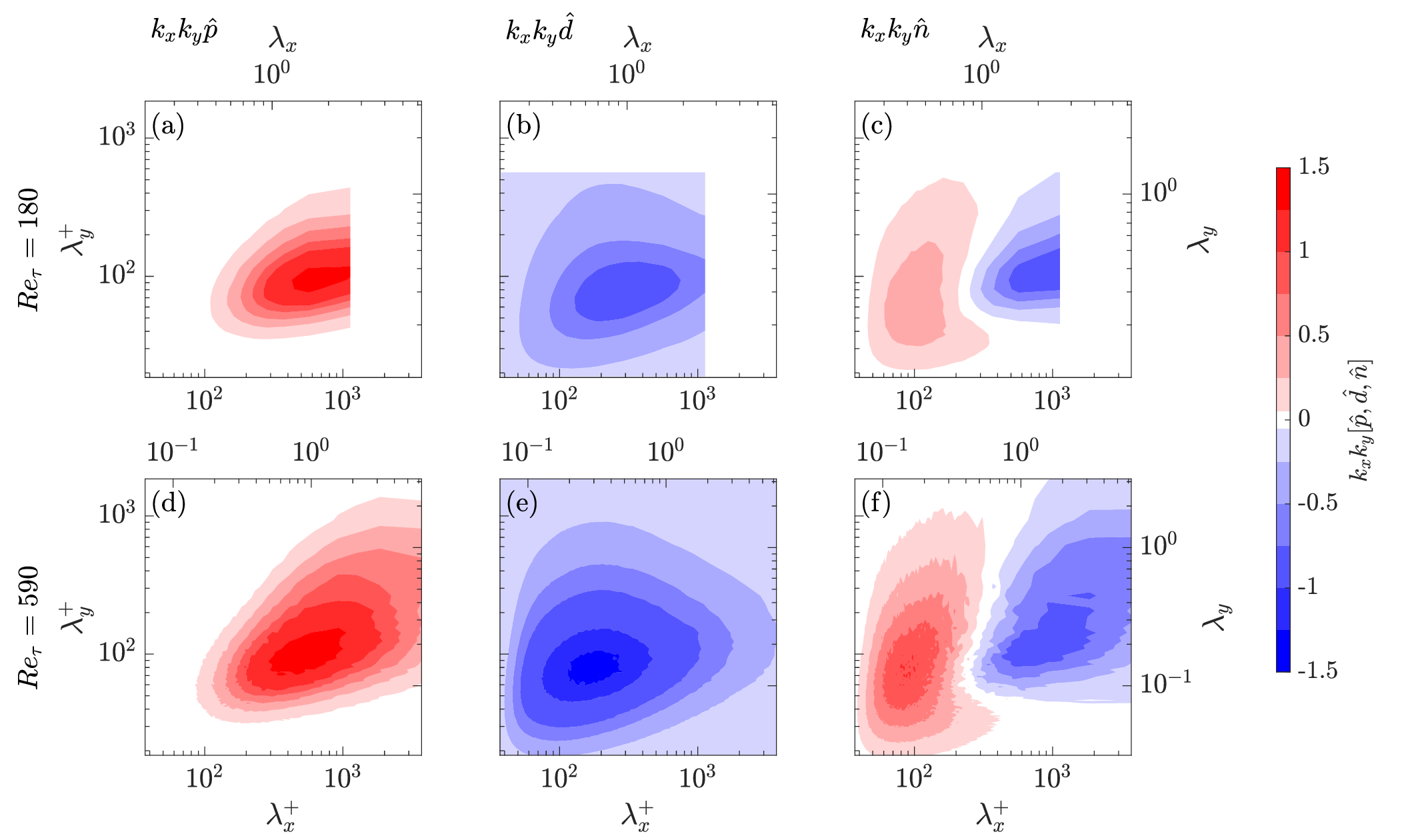}
\caption{
(a,d) Premultiplied production spectra $k_x k_y \hat{p}$; (b,e) premultiplied (negative) dissipation spectra $k_x k_y \hat{d}$; (c,f) premultiplied nonlinear energy transfer spectra $k_x k_y \hat{n}$.
(a,b,c), $Re_{\tau}=180$; (d,e,f), $Re_{\tau}=590$.}
\label{fig_PDN_log}
\end{figure}

This section revisits the wall-normal integrated energy transfer for a single mode, as stated in equation \eqref{eqa_PDN} \citep{symon2021energy}.
Different from \citet{symon2021energy}, we use two full channel datasets, and visualise the premultiplied energy transfer spectra, as shown in figure \ref{fig_PDN_log}. 
The relationship between wavenumber and wavelength is $\lambda = 2\pi/k$, where $\lambda$ refers to wavelength and $k$ refers to wavenumber. 
As for the production spectra $\hat{p}$ and dissipation spectra $\hat{d}$ (figures \ref{fig_PDN_log}(a,b,d,e)), we see that the wall-normal integrated energy transfer spectra show relatively $Re_{\tau}$-independent features at these two Reynolds numbers: their peaks are nearly aligned. 
The locations of the peaks are in line with the previous study that the production peak occurs at $\lambda_x^+ \approx 600$, $\lambda_y^+ \approx 100$ (at $z^+ \approx 15$) and the peak for dissipation caused by streamwise velocities occurs at $\lambda_x^+ \approx 200$, $\lambda_y^+ \approx 70$ (at $z^+ \approx 70$) \citep{lee2019spectral}. 
The Reynolds-Orr equation states that the energy source for turbulence is production and the energy sink is dissipation \citep{schmid2002stability}. 
However, the energy source and sink are characterised by different structures, because the production and dissipation spectra are not identical as seen from figures \ref{fig_PDN_log}(a,b,d,e).
When viewed in terms of individual Fourier modes, equation \eqref{eqa_PDN} states that the gap between production and dissipation is bridged by nonlinear energy transfer which is conservative as shown in equation \eqref{eqa_N_0}.
As for the nonlinear energy transfer spectra $\hat{n}$ (figures \ref{fig_PDN_log}(c,f)), we see the same streamwise forward cascade in which energy is transferred from large streamwise scales to small streamwise scales at $Re_{\tau}=180$ and $590$. 

Recall that $\hat{n}$ represents the net energy one mode receives from all other modes through nonlinear interactions.
Observing figures \ref{fig_PDN_log}(c,f), we see that there is a band of modes near $\lambda_x^+ \approx 300$ with near-zero net energy transfer $\hat{n}(k_x,k_y) \approx 0$. 
However, $\hat{n}=0$ alone cannot distinguish these two possible cases: first, these modes do not participate in nonlinear interactions; second, these modes gain and lose approximately the same amount of energy resulting in a near-zero energy transfer. 
Similarly, for modes for which $\hat{n} \neq 0$, $\hat{n}$ only provides the net energy transfer for one mode without giving the detailed budget. 
We see the same streamwise forward cascade at $Re_{\tau}=180$ and $590$ (figures \ref{fig_PDN_log}(c,f)) by examining $\hat{n}$.
However, it is not the only piece of information we can obtain from the nonlinear interactions.
In order to explore nonlinear interactions in more detail, we investigate the mode-to-mode nonlinear energy transfer using the variable $\hat{M}_{(s_x,s_y)(k_x,k_y)}$ defined in equation \eqref{eqa_M}.

\subsection{Energy transfer pathways} \label{section_3_examples}
We first use two examples to interpret the introduced four-dimensional variable $\hat{M}_{(s_x,s_y)(k_x,k_y)}$ which represents energy transferred nonlinearly from mode $(s_x,s_y)$ to mode $(k_x,k_y)$, as shown in equation \eqref{eqa_M} and Appendix \ref{appendix_B}.
For each example, we first choose a streamwise wavenumber and a spanwise wavenumber for mode $(k_x,k_y)$, which remains fixed. 
Then, we vary the streamwise and spanwise wavenumbers for mode $(s_x,s_y)$.
Both examples use the $Re_{\tau}=180$ dataset.
First, $\hat{M}$ is visualised with linear axes because modes containing zero wavenumbers cannot be shown on a premultiplied energy spectrum. 
Second, the premultiplied spectrum $s_x s_y \hat{m}$ is shown.

\begin{figure}
\centering
\includegraphics[scale=0.40]{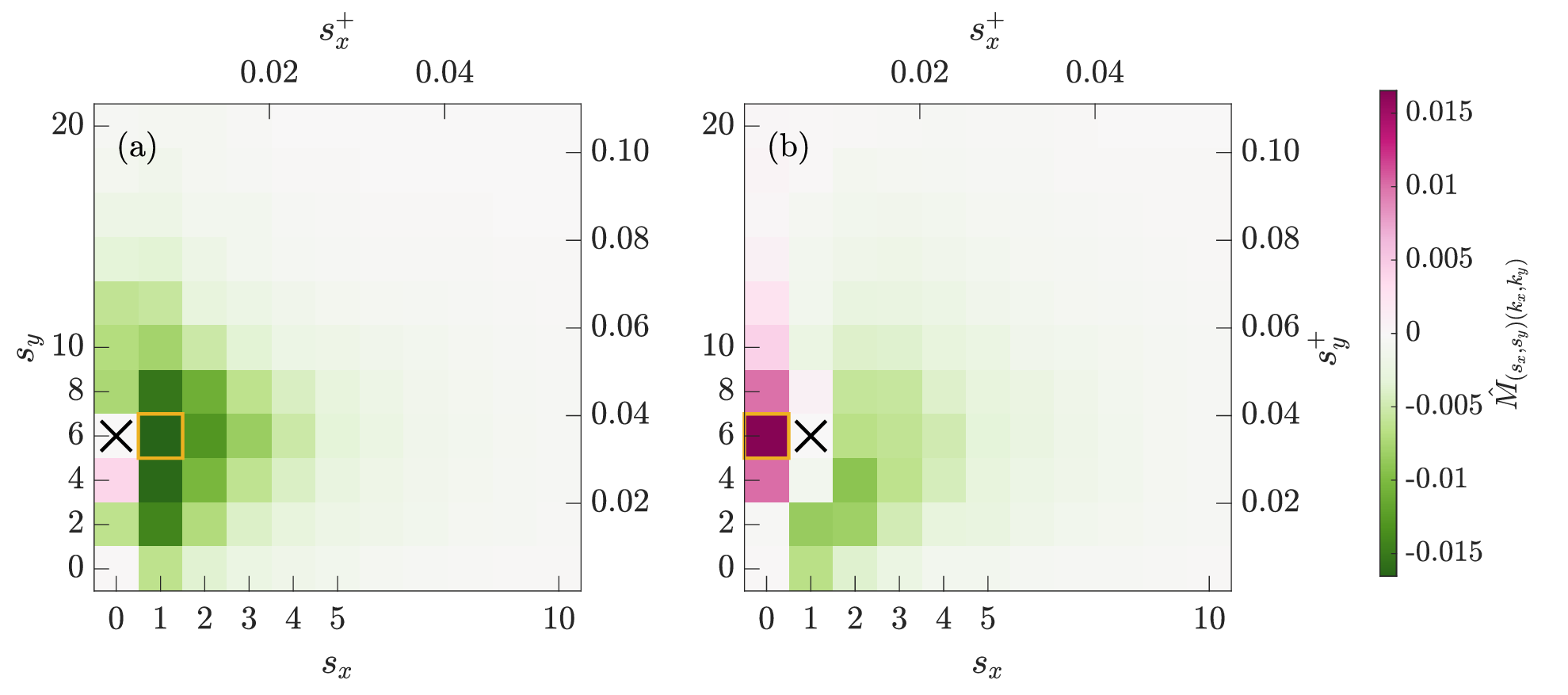}
\caption{
(a) $\hat{M}_{(s_x,s_y)(0,6)}$, the black cross marks the fixed mode $(0,6)$; (b) $\hat{M}_{(s_x,s_y)(1,6)}$, the black cross marks the fixed mode $(1,6)$. Modes marked in orange boxes are used to illustrate the property stated in equation \eqref{eqa_energy_transfer_mechanism_opposite}. 
The data is calculated for the $Re_{\tau}=180$ case. 
 }
\label{fig_M_example}
\end{figure}

\begin{figure}
\centering
\includegraphics[scale=0.40]{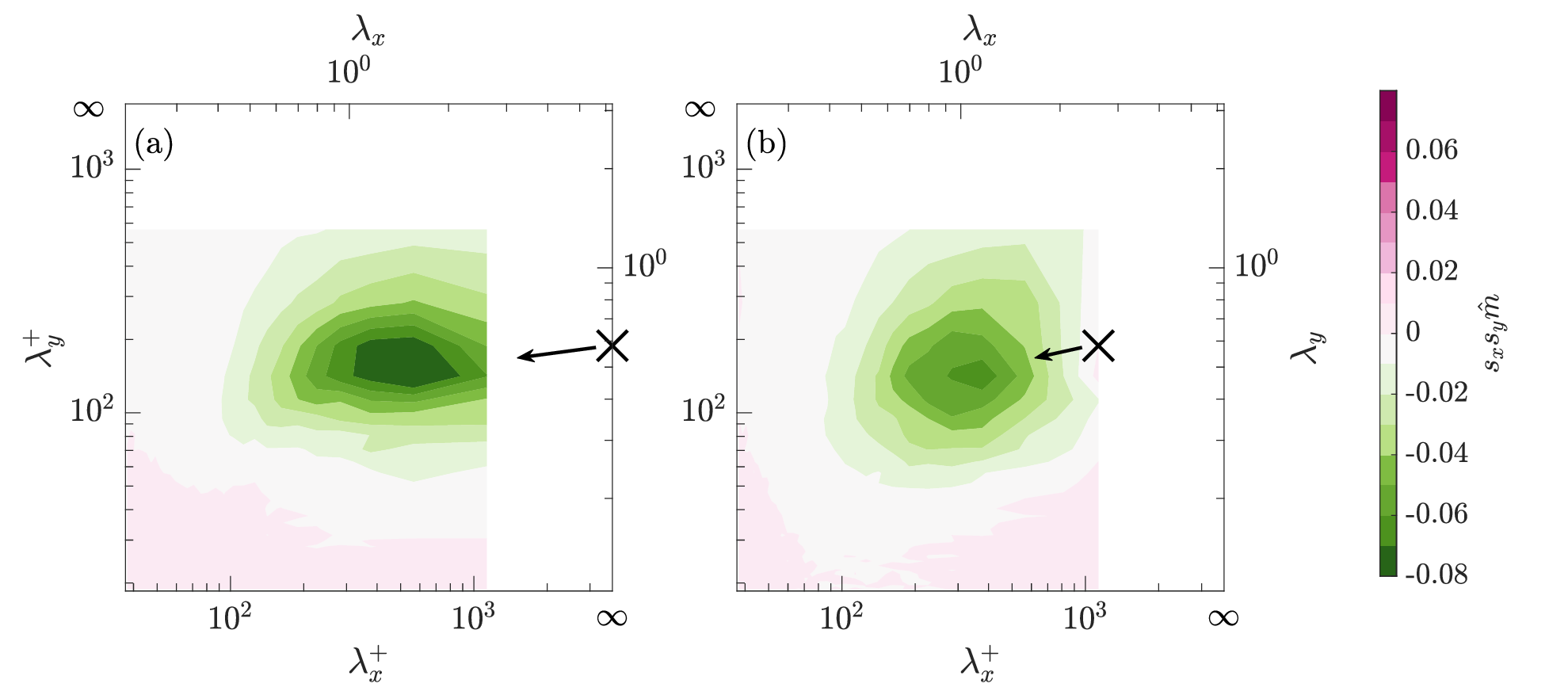}
\caption{
(a) The premultiplied spectrum $s_x s_y \hat{m}_{(s_x,s_y)(0,6)}$, the black cross marks the fixed mode $(0,6)$ corresponding to $(\lambda_x^+,\lambda_y^+)=(\infty,188)$.
(b) The premultiplied spectrum $s_x s_y \hat{m}_{(s_x,s_y)(1,6)}$, the black cross marks the fixed mode $(1,6)$ corresponding to $(\lambda_x^+,\lambda_y^+)=(1130,188)$.
The arrow in each figure marks the dominant energy transfer direction. 
 }
\label{fig_M_spectrum_example}
\end{figure}

For the first example, we choose to fix mode $(0,6)$ corresponding to $(\lambda_x^+ = \infty, \lambda_y^+=188)$.
This mode has significant production at $Re_{\tau}=180$ which means that this mode gains significant energy from the mean flow.
We would like to understand how this streamwise-constant mode redistributes energy to other modes. 
Figure \ref{fig_M_example}(a) shows $\hat{M}_{(s_x,s_y)(0,6)}$ which quantifies the energy that mode $(0,6)$ marked by the black cross gains from modes in pink and loses to modes in green.  
We see that apart from gaining energy from the wider mode $(0,4)$ (wider in the spanwise direction), it loses energy to all other modes, including the even wider mode $(0,2)$. 
The colour intensity tells us that local nonlinear energy transfer between mode $(0,6)$ and its neighbouring modes in Fourier space is strong \citep{domaradzki1994energy, brasseur1994interscale, cho2018scale}. 
In particular, mode $(0,6)$ loses the most energy to the next smallest streamwise scales with $s_x=1$ outlined in orange, representing a forward energy cascade. 
This forward cascade can also be observed from the arrow in the premultiplied spectrum in figure \ref{fig_M_spectrum_example}(a).

For the second example, we choose to fix mode $(1,6)$ corresponding to $(\lambda_x^+ = 1130, \lambda_y^+=188)$.
Figure \ref{fig_M_example}(b) shows $\hat{M}_{(s_x,s_y)(1,6)}$.
From the first example, we see that mode $(0,6)$ loses the most energy to this mode $(1,6)$. 
The two modes highlighted in orange boxes in figures \ref{fig_M_example}(a,b) have the same magnitude but opposite signs, respecting equation \eqref{eqa_energy_transfer_mechanism_opposite}.
We see the similar forward energy cascade in which mode $(1,6)$ gains energy from the next largest streamwise scale with $s_x=0$ and loses energy to the next smallest streamwise scale with $s_x=2$ but larger spanwise scale with $s_y=4$.
In general, inspecting the spanwise wavenumbers, we observe that mode $(1,6)$ loses a significant amount of energy to scales with larger spanwise wavelengths $(s_y<6)$.
This corresponds to a spanwise inverse energy cascade in which energy is transferred from small spanwise scales to large spanwise scales \citep{cimarelli2013paths,cimarelli2016cascades,cho2018scale}, though this inverse spanwise energy cascade is not obvious from the arrow in the premultiplied spectrum in figure \ref{fig_M_spectrum_example}(b).

\begin{figure}
\centering
\includegraphics[scale=0.40]{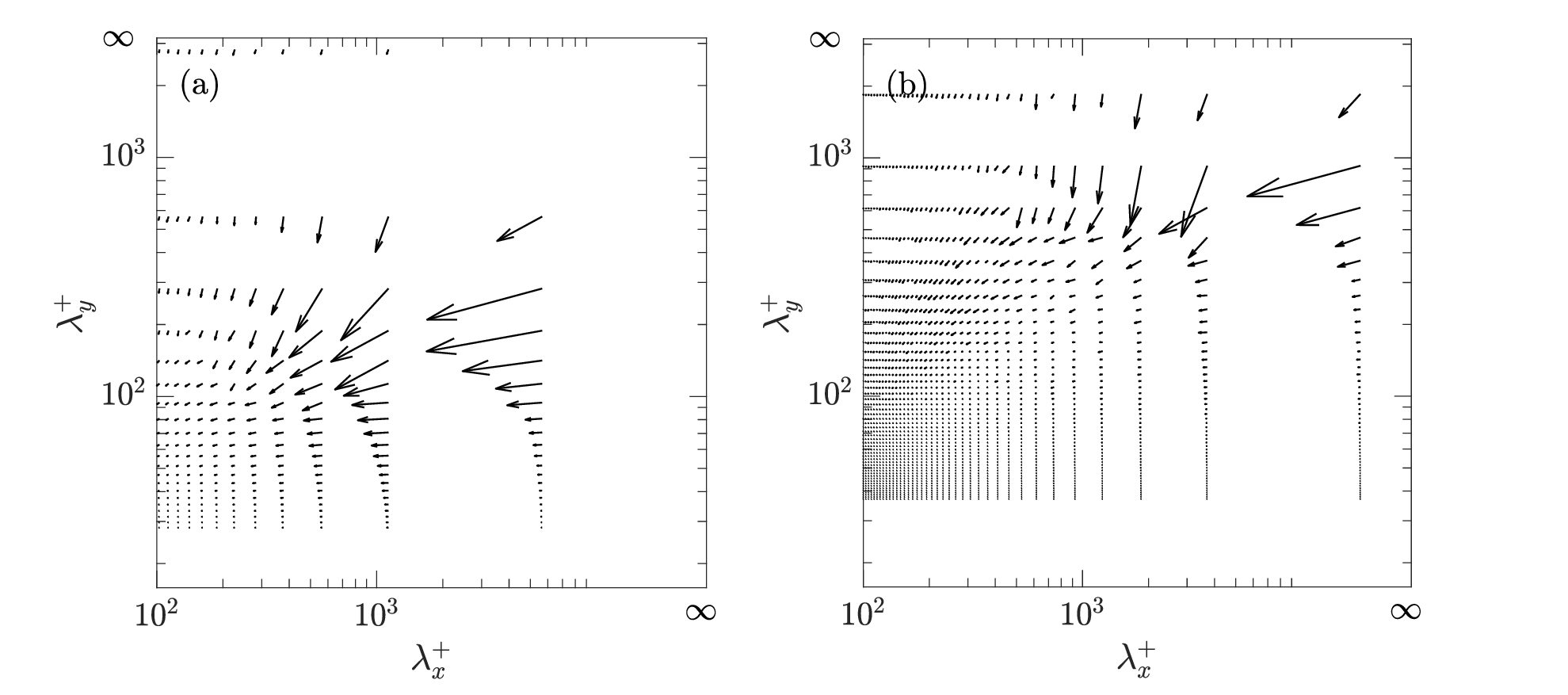}
\caption{
Dominant energy transfer pathways. 
(a) $Re_{\tau}=180$;
(b) $Re_{\tau}=590$.
 }
\label{fig_M_pathway_180_590}
\end{figure}

For each example, we obtain one dominant energy transfer pathway from the fixed mode $(k_x,k_y)$ to the mode to which the fixed mode $(k_x,k_y)$ loses the most energy (as shown by the arrows in figure \ref{fig_M_spectrum_example}).
We can further change the fixed mode $(k_x,k_y)$ to obtain more dominant energy transfer pathways in order to better visualise the nonlinear energy transfer in streamwise-spanwise wavenumber space, as shown in figure \ref{fig_M_pathway_180_590}.
For both $Re_{\tau}=180$ and $590$, we see that there are large left-pointing arrows corresponding to the forward streamwise energy cascade and large bottom-pointing arrows corresponding to the forward spanwise energy cascade. 
It should be noted that each arrow in figure \ref{fig_M_pathway_180_590} only illustrates the dominant energy transfer pathway for each fixed $(k_x,k_y)$ case without showing other less dominant energy transfer pathways.
In addition, the energy transfer pathways describe only the statistical properties because the variable $\hat{M}$ is a time-averaged quantity \eqref{eqa_M}. 

The two examples (figure \ref{fig_M_example}) illustrate the quantification of mode-to-mode nonlinear energy transfer in streamwise-spanwise wavenumber space. 
For mode $(0,6)$, we can further calculate how much energy mode $(0,6)$ gains in total due to nonlinear interactions by summing all the modes in pink in figure \ref{fig_M_example}(a). 
Similarly, we can calculate how much energy mode $(0,6)$ loses in total due to nonlinear interactions by summing all the modes in green in figure \ref{fig_M_example}(a). 
The next section aims to calculate the net energy transfer gain and loss due to nonlinear interactions for each mode in streamwise-spanwise wavenumber space. 

\subsection{Decomposition of net nonlinear energy transfer $\hat{N}$}
\label{section_source_recipient}
With the introduced variable $\hat{M}$ \eqref{eqa_N_convolution}, we can decompose $\hat{N}$ for a given mode $(k_x,k_y)$ into positive and negative contributions:
\begin{subequations}
\begin{align}
\hat{N}^{+}(k_x,k_y) &= \int_{0}^{\infty} \int_{0}^{\infty} \hat{m}_{(s_x,s_y)(k_x,k_y)} \{ \hat{m}>0 \} \; \ddd s_x \ddd s_y  \label{eqa_N_positive} \\
\hat{N}^{-}(k_x,k_y) &= \int_{0}^{\infty} \int_{0}^{\infty} \hat{m}_{(s_x,s_y)(k_x,k_y)} \{ \hat{m}<0 \} \; \ddd s_x \ddd s_y  \label{eqa_N_negative}%
\end{align}
\end{subequations}
where $\{\;\}$ is an indicator. 
An indicator is equal to $1$ when the argument is true and $0$ when the argument is false. 
The positive energy transfer $\hat{N}^+$ quantifies the total energy mode $(k_x,k_y)$ gains from other modes.
The negative energy transfer $\hat{N}^-$ quantifies the total energy mode $(k_x,k_y)$ loses to other modes.
According to equation \eqref{eqa_N_convolution}, these two variables are linked to the net energy transfer $\hat{N}$ by
\begin{equation}
    \hat{N}(k_x,k_y)     = \hat{N}^{+}(k_x,k_y) + \hat{N}^{-}(k_x,k_y)
    \label{eqa_N_Npositive_Nnegative}
\end{equation}
Although seeming straightforward, this decomposition cannot be attained without formulating the mode-to-mode nonlinear energy transfer $\hat{M}_{(s_x,s_y)(k_x,k_y)}$. 

\begin{figure}
\centering
\includegraphics[scale=0.40]{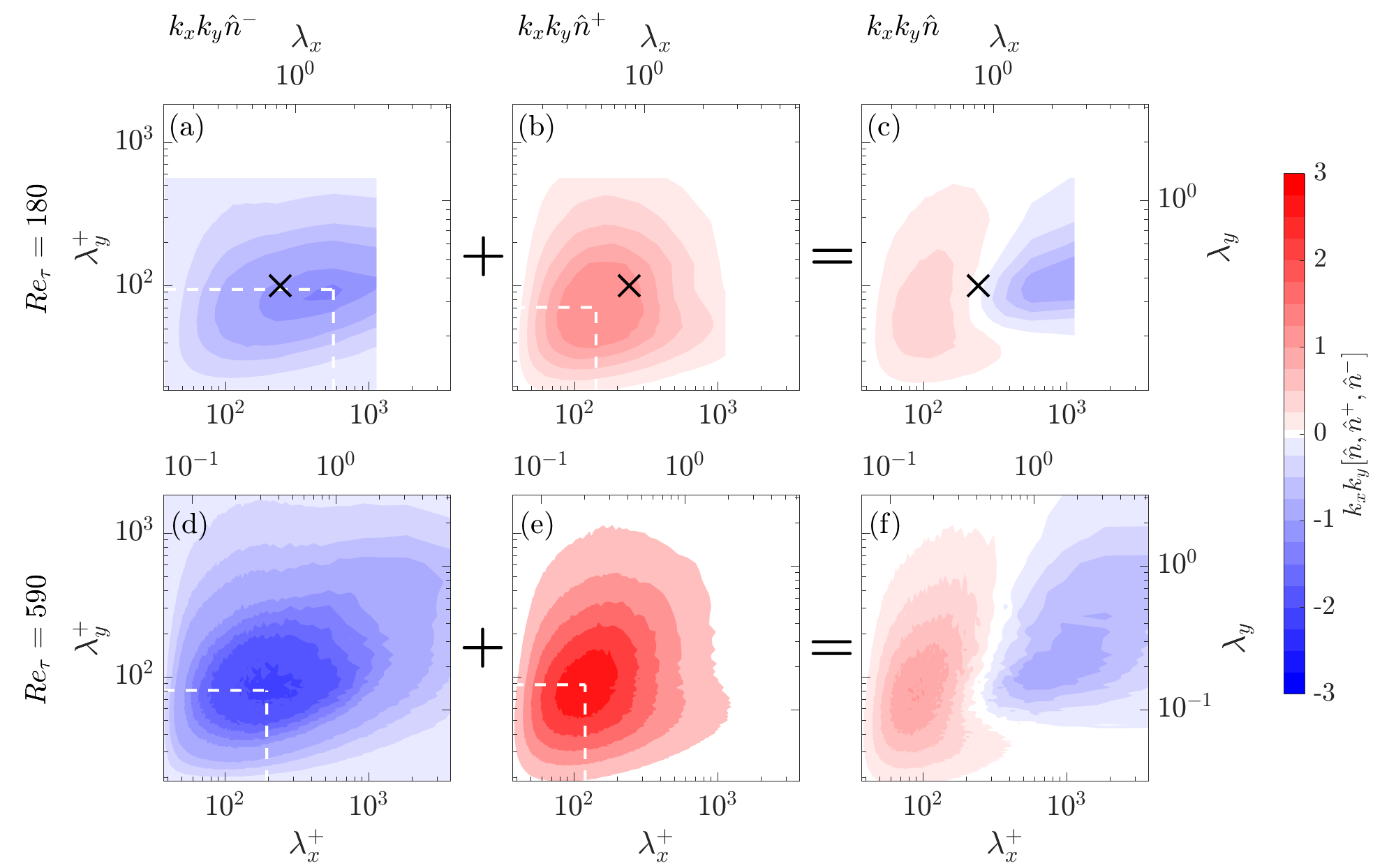}
\caption{Decomposition of the net nonlinear energy transfer: $\hat{n}^- + \hat{n}^+ = \hat{n}$. 
(a,d) Premultiplied negative nonlinear energy transfer spectra $k_x k_y \hat{n}^-$, dashed lines mark the peak; 
(b,e) Premultiplied positive nonlinear energy transfer spectra $k_x k_y \hat{n}^+$, dashed lines mark the peak; 
(c,f) Premultiplied net nonlinear energy transfer spectra $k_x k_y \hat{n}$. 
(a,b,c), $Re_{\tau}=180$; (d,e,f), $Re_{\tau}=590$. 
The black crosses in (a,b,c) mark a mode with $\hat{n} \approx 0$ for explanation purposes. }
\label{fig_N_positive_N_negative_log180590}
\end{figure}

Figures \ref{fig_N_positive_N_negative_log180590}(a)(b) quantify the energy transfer gain and loss for each mode at $Re_{\tau} = 180$. 
We see the dual characteristic of nonlinear energy transfer that each mode acts as an energy source and energy recipient.
The difference between the energy transfer gain and loss for each mode results in the net nonlinear energy transfer spectrum $\hat{n}$, as shown in figure \ref{fig_N_positive_N_negative_log180590}(c) and respecting equation \eqref{eqa_N_Npositive_Nnegative}. 
Figure \ref{fig_N_positive_N_negative_log180590}(c) is the same as figure \ref{fig_PDN_log}(c). 
Here, we address one consideration raised at the end of \S\ref{section_PDN}.
We see that the mode marked by the black cross where $\hat{n} \approx 0$ in figure \ref{fig_N_positive_N_negative_log180590}(c) has non-negligible net energy loss and net energy gain as observed in the corresponding locations in figures \ref{fig_N_positive_N_negative_log180590}(a)(b). 
This indicates that this mode gains and loses approximately equal amount of energy, resulting in near-zero net nonlinear energy transfer. 
It would be incorrect to infer that this mode does not participate in nonlinear interactions purely judging from the net nonlinear energy transfer spectrum $\hat{n}$. 

Now we answer the other question raised at the end of \S\ref{section_PDN}.
As mentioned previously, the net energy transfer spectra $\hat{n}$ reveal the same streamwise forward cascade at $Re_{\tau}=180$ and $590$ (figures \ref{fig_N_positive_N_negative_log180590}(c,f)). 
At $Re_{\tau}=180$, we see that the negative energy transfer spectrum $\hat{n}^-$ peak is at $\lambda_x^+ \approx 600,\; \lambda_y^+ \approx 100$ and the positive energy transfer spectrum $\hat{n}^+$ peak is at $\lambda_x^+ \approx 150,\; \lambda_y^+ \approx 70$ indicated by the white dashed lines in figures \ref{fig_N_positive_N_negative_log180590}(a,b). 
This indicates that large streamwise scales lose the most energy and small streamwise scales gain the most energy.
Thus, figures \ref{fig_N_positive_N_negative_log180590}(a,b) illustrate the streamwise forward cascade, which aligns with the net energy transfer spectrum (figure \ref{fig_N_positive_N_negative_log180590}(c)).

At $Re_{\tau}=590$, we see that the negative energy transfer spectrum $\hat{n}^-$ peak is at $\lambda_x^+ \approx 200,\; \lambda_y^+ \approx 90$ and the positive energy transfer spectrum $\hat{n}^+$ peak is at $\lambda_x^+ \approx 150,\; \lambda_y^+ \approx 80$ indicated by the white dashed lines in figures \ref{fig_N_positive_N_negative_log180590}(d,e). 
From $Re_{\tau}=180$ to $590$, there is a significant negative energy transfer spectrum peak shift to a much smaller $\lambda_x^+$.
Figures \ref{fig_N_positive_N_negative_log180590}(d,e) reveal that modes losing the most energy and modes gaining the most energy are of similar sizes, as the two peaks are near to each other.
This is a new piece of information obtained from $\hat{n}^+$ and $\hat{n}^-$ as this wavenumber area responsible for the significant energy gain and loss cannot be seen from the net energy transfer spectrum (figure \ref{fig_N_positive_N_negative_log180590}(f)).
Note that the above discussions are for the whole channel since $\hat{M}$ is wall-normal integrated nonlinear energy transfer which hides the nonlinear energy transfer in the wall-normal direction. 

\subsection{Nonlinear energy transfer of three structures} \label{section_three_structures}
Linear analysis explains the energy amplification mechanisms of different structures \citep{schmid2002stability, jovanovic2005componentwise}, leaving the nonlinear part relatively unexplored. 
Since nonlinear energy transfer is conservative, the nonlinear energy transfer among different structures should be interpreted as energy redistribution. 
In this section, we use $\hat{M}$ to investigate the energy redistribution of three different structures: streamwise streaks, oblique waves and Tollmien–Schlichting (TS) waves.
Since streamwise streaks and TS waves are characterised by zero wavenumbers which cannot be shown on premultiplied energy transfer spectra using logarithmic axes, we first present the nonlinear energy transfer using linear axes. 
It should be noted that the original definition of TS waves refers to the velocity fluctuations based on the laminar profile. 
Here, the velocity fluctuations are based on the turbulent mean profile. 

Following \citet{jovanovic2005componentwise}, streamwise streaks are characterised by $k_x \approx 0, k_y \approx O(1)$.
Therefore, we define streamwise streaks as modes with $k_x=0$ $(\lambda_x^+ = \infty)$. 
We define the nonlinear energy transfer of streamwise streaks as:
\begin{equation}
    \hat{M}_{(s_x,s_y) \to streaks}, \; k_{x(streaks)}=0\;(\lambda_{x(streaks)}^+ = \infty)
    \label{M_streaks}
\end{equation}
Oblique waves are characterised by $k_x \approx O(1), k_y \approx O(1)$.
Considering the geometry of the channel box used in this study, we define oblique waves as modes satisfying $k_y=2k_x$ $(\lambda_x^+ = 2\lambda_y^+)$. 
We define the nonlinear energy transfer of oblique waves as:
\begin{equation}
    \hat{M}_{(s_x,s_y) \to obwaves}, \; k_{y(obwaves)}=2k_{x(obwaves)}\;(\lambda_{x(obwaves)}^+ = 2\lambda_{y(obwaves)}^+)
    \label{M_obwaves}
\end{equation}
TS waves are characterised by $k_x \approx O(1), k_y \approx 0$.
Therefore, we define TS waves as modes with $k_y=0$ $(\lambda_y^+ = \infty)$. 
We define the nonlinear energy transfer of TS waves as:
\begin{equation}
    \hat{M}_{(s_x,s_y) \to TSwaves}, \; k_{y(TSwaves)}=0 \;(\lambda_{y(TSwaves)}^+ = \infty)
    \label{M_TSwaves}
\end{equation}
Note that the purpose of this section is to discuss the above three distinct structures rather than discuss all modes in streamwise-spanwise wavenumber space.

\begin{figure}
\centering
\includegraphics[scale=0.40]{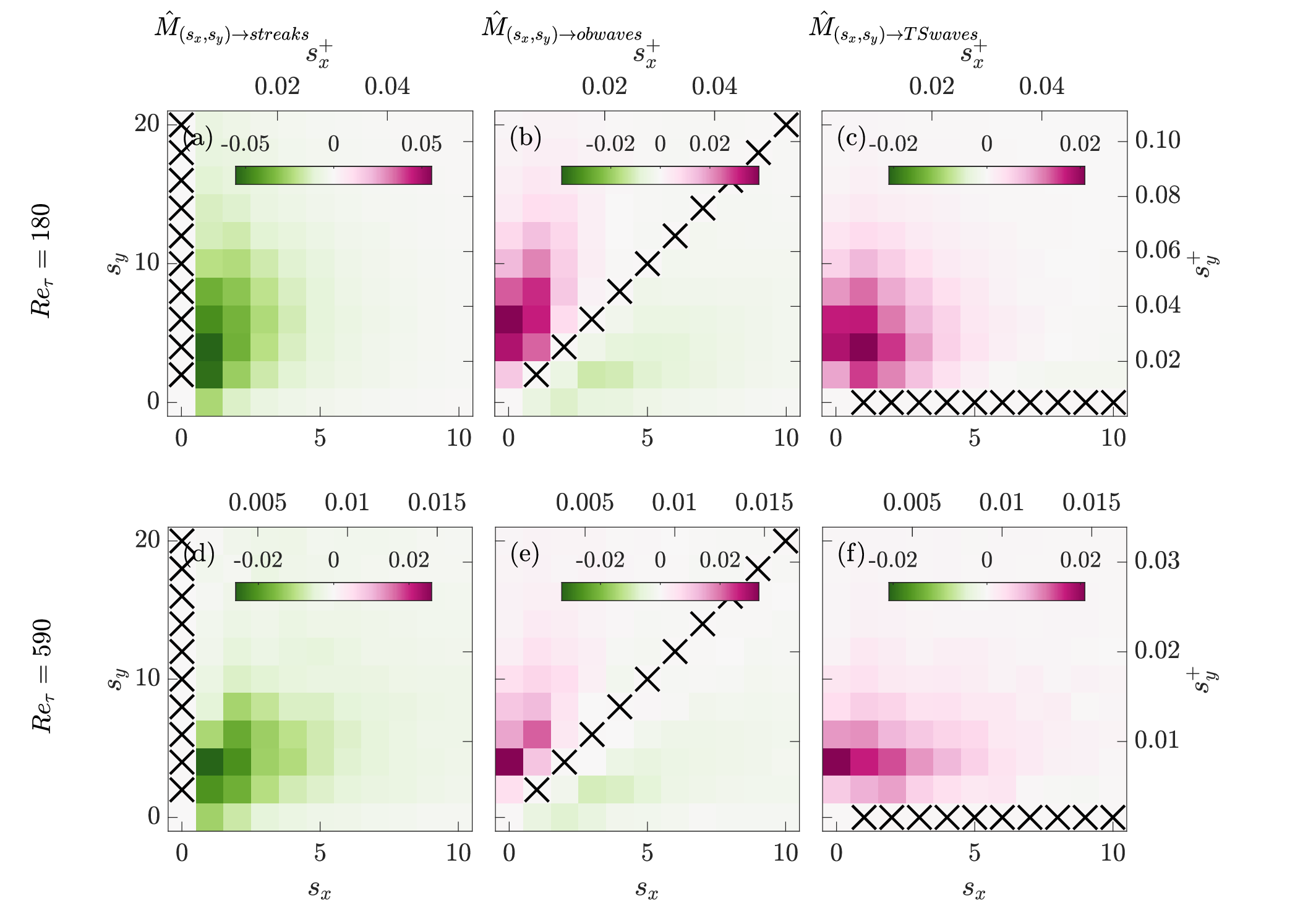}
\caption[Nonlinear energy transfer of streamwise streaks, oblique waves and TS waves at $Re_{\tau}=180$ and $590$.]
{(a,d) $\hat{M}_{(s_x,s_y) \to streaks}$; (b,e) $\hat{M}_{(s_x,s_y) \to obwaves}$; (c,f) $\hat{M}_{(s_x,s_y) \to TSwaves}$ visualised using discrete modes. (a,b,c), $Re_{\tau} = 180$; (d,e,f), $Re_{\tau} = 590$. Black crosses mark the modes of the investigated structures.}
\label{fig_M_structures180590}
\end{figure}

Figure \ref{fig_M_structures180590} shows the nonlinear energy transfer of the three different structures at $Re_{\tau}=180$ and $590$. 
As for the interpretation of each subplot, the specific structure (indicated by the black crosses) gains energy from modes in pink and loses energy to modes in green. 
We see that streamwise streaks (figures \ref{fig_M_structures180590}(a,d)) lose energy to smaller streamwise scales ($s_x>0$), exhibiting a streamwise forward cascade. 
This could be linked to streak breakdown which is one phase of the self-sustaining process (SSP). 
Due to the instability of long streamwise streaks, they break down into smaller streamwise streaks \citep{hamilton1995regeneration}. 
For oblique waves (figures \ref{fig_M_structures180590}(b,e)), we see that they generally gain energy from scales with larger aspect ratios $ s_y / s_x$ (which is equivalent to $k_y/k_x$) \citep{symon2021energy} and lose energy to scales with smaller aspect ratios, exhibiting a transverse cascade \citep{lee2019spectral, symon2021energy}. 
For TS waves (figures \ref{fig_M_structures180590}(c,f)), we see that they mainly gain energy from smaller spanwise scales, exhibiting a spanwise inverse cascade \citep{cimarelli2013paths,cimarelli2016cascades,cho2018scale}. 
If we think of streamwise streaks with infinite aspect ratio $s_y / s_x \to \infty$ and TS waves with zero aspect ratio $s_y / s_x \to 0$, then we could conclude that there exists energy transfer from scales with large aspect ratio to scales with small aspect ratio. 
In terms of the shapes of scales, this transverse cascade refers to energy transfer from streamwise-elongated scales to spanwise-elongated scales. 
It is also worth noting that the transverse cascade of the three structures is not substantially influenced by the Reynolds number for this range. 

\begin{figure}
\centering
\includegraphics[scale=0.35]{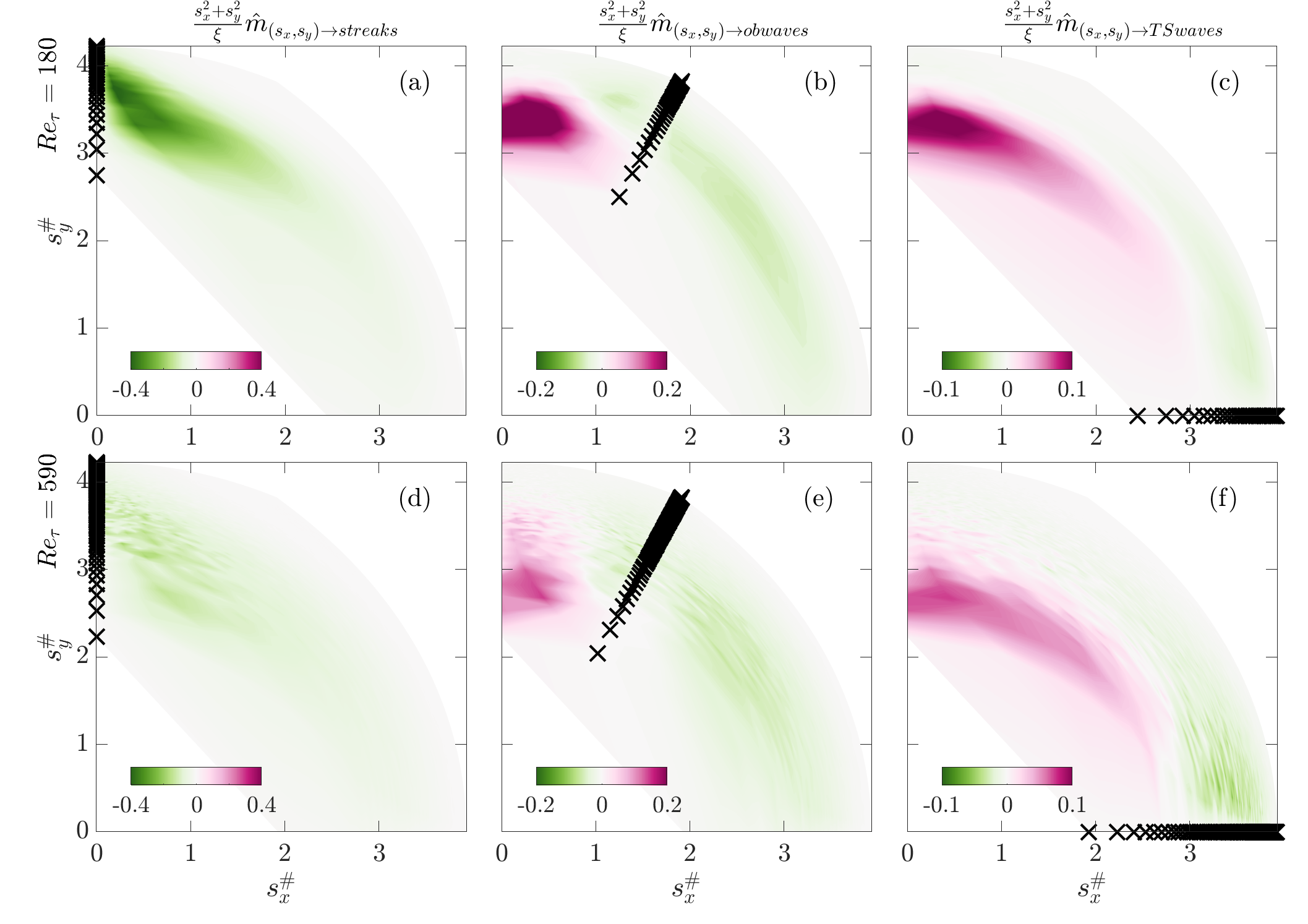}
\caption[Premultiplied nonlinear energy transfer spectra of streamwise streaks, oblique waves and TS waves in log-polar coordinate at $Re_{\tau}=180$ and $590$.]
{Log-polar premultipled energy spectra (a,d) $\frac{s_x^2+s_y^2}{\xi} \hat{m}_{(s_x,s_y) \to streaks}$; (b,e) $\frac{s_x^2+s_y^2}{\xi} \hat{m}_{(s_x,s_y) \to obwaves}$; (c,f) $\frac{s_x^2+s_y^2}{\xi} \hat{m}_{(s_x,s_y) \to TSwaves}$. (a,b,c), $Re_{\tau} = 180$; (d,e,f), $Re_{\tau} = 590$. Black crosses mark the modes of the investigated structures.}
\label{fig_M_structures_logpolar_180590}
\end{figure}

The above discussion tells us that the transverse energy cascade is related to $s_y / s_x$ (which is equivalent to $k_y/k_x$).
We can also visualise the nonlinear energy transfer spectrum on a log-polar coordinate in which $s_y / s_x$ corresponds to a certain slope \citep{lee2019spectral}, as shown in figure \ref{fig_M_structures_logpolar_180590}.
Following \citet{lee2019spectral}, $s_x^{\#} = \xi s_x / \sqrt{s_x^2+s_y^2}$, $s_y^{\#} = \xi s_y / \sqrt{s_x^2+s_y^2}$ and $\xi = \log(\sqrt{s_x^2+s_y^2}/k_{ref})$ with $k_{ref}=50000$.
From figure \ref{fig_M_structures_logpolar_180590}(a,d), the streamwise streaks located on the $a^{\#}=0$ axis lose energy to other modes, corresponding to the finding that modes located on the $s_x^{\#}=0$ axis have significant net negative nonlinear energy transfer \citep{lee2019spectral}.
From figures \ref{fig_M_structures_logpolar_180590}(b,e), the oblique waves located on the $s_y^{\#} = 2 s_x^{\#}$ gain energy from modes satisfying $s_y^{\#} > 2 s_x^{\#}$ and lose energy to modes satisfying $s_y^{\#} < 2 s
_x^{\#}$, corresponding to the transverse cascade mentioned previously. 
From figures \ref{fig_M_structures_logpolar_180590}(c,f), the TS waves located on the $s_y^{\#}=0$ axis mainly gain energy but also lose a small amount of energy. 
We see that one advantage of using log-polar premultiplied spectra over the traditional premultiplied spectra is that we can visualise modes which contain one zero wavenumber (either $s_x=0$ or $s_y=0$).

Note that the above discussion only concerns the nonlinear energy transfer as one component in the energy transfer balance \eqref{eqa_PDN}. 
One should not interpret that energy merely originates from streamwise streaks, goes through oblique waves and finally dissipates at TS waves. 
Equation \eqref{eqa_PDN} gives the energy transfer budget for each mode. 
To evaluate the overall energy transfer balance comprehensively, we need to consider the production and dissipation as well. 
Motivated by \S\ref{section_source_recipient}, we decompose the net nonlinear energy transfer of each structure into its positive and negative parts.
According to equations \eqref{eqa_PDN} and \eqref{eqa_N_Npositive_Nnegative}, for each structure, the energy transfer budget is
\begin{equation}
\hat{P} + \hat{D} + \hat{N}^+ + \hat{N}^- = 0
\label{eqa_energy_balance_structure}
\end{equation}

\begin{figure}
\centering
\includegraphics[scale=0.40]{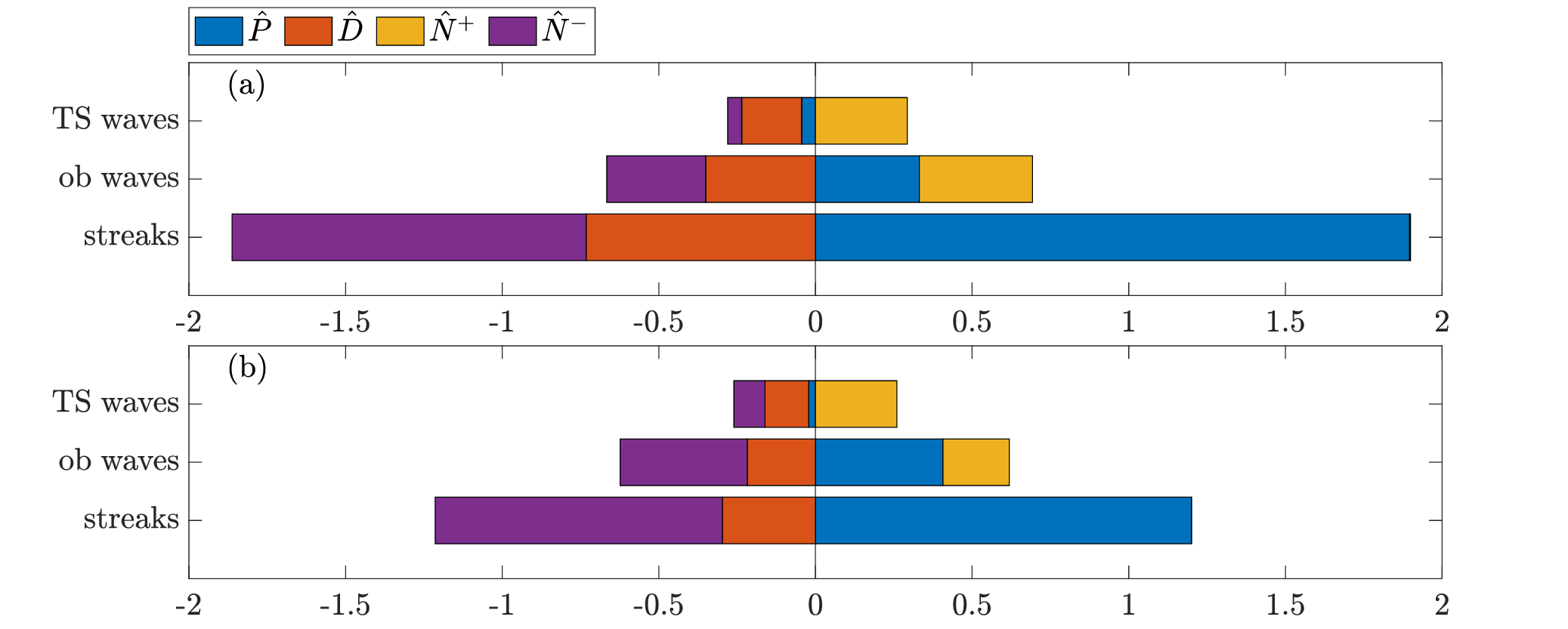}
\caption[Energy budgets for the streamwise streaks, oblique waves and TS waves at $Re_{\tau}=180$ and $590$.]
{ 
Energy budgets for the streamwise streaks, oblique waves and TS waves.
(a) $Re_{\tau}=180$; (b) $Re_{\tau}=590$.}
\label{fig_M_structures_energy_transfer_180590}
\end{figure}

The energy budgets \eqref{eqa_PDN} for the three structures are shown in figure \ref{fig_M_structures_energy_transfer_180590}.
For the streamwise streaks, they receive a large amount of energy from the mean flow through production as seen from the blue bars and they lose a large amount of energy nonlinearly to other modes as seen from the purple bars. 
For the oblique waves, apart from gaining energy from the mean flow and dissipating energy, they gain and lose energy nonlinearly through interacting with other modes as seen from the purple and yellow bars. 
For the TS waves, the only energy gain is through nonlinear energy transfer as seen from the yellow bars. 
Streamwise streaks and oblique waves gain energy from the mean flow (positive production), while TS waves lose energy to the mean flow (negative production).
Dissipation occurs for all three structures as seen from the red bars.
Although the values are different at the two Reynolds numbers, different types of energy transfer show similar trends in the three structures. 

\color{black}

\subsection{Forward cascade and inverse cascade} \label{section_forward_inverse_cascade}
In this section, we use $\hat{M}$ to quantify the forward and inverse cascades between the large scales and small scales in the spirit of large-eddy simulation (LES). 
For channel flows, resolved scales could be set by modes belonging to a rectangular region determined by the conditions of $k_x \leq k_{xC}$ and $k_y \leq k_{yC}$, where $k_{xC}$ and $k_{yC}$ are cutoff wavenumbers \citep{germano1991dynamic, hartel1994subgrid, domaradzki1994energy}. 
These cutoff wavenumbers are determined by the choice of a single variable $n_C$:
\begin{subequations}\label{eqa_nC}
\begin{gather}
k_{xC} = \frac{2\pi n_C}{L_x},   \quad   k_{yC} = \frac{2\pi n_C}{L_y}     \tag{\theequation a,b}
\end{gather}
\end{subequations}

\begin{figure}
\centering
\includegraphics[scale=0.35]{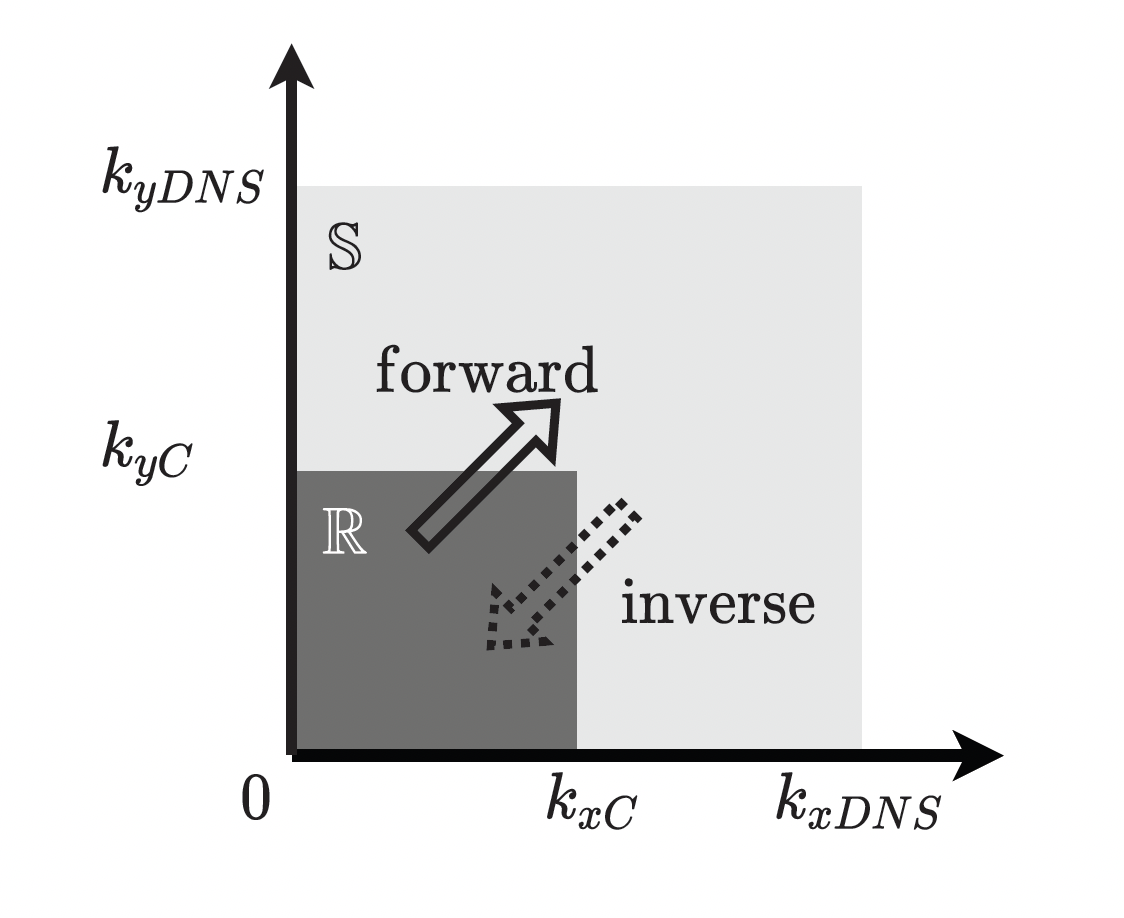}
\caption[A sketch illustrating the forward cascade and inverse cascade between the resolved scales and subgrid scales.]
{A sketch illustrating the forward cascade and inverse cascade between the resolved-scale region $\mathbb{R}$ marked in dark grey and the subgrid-scale region $\mathbb{S}$ marked in light grey. 
Region $\mathbb{R}$ contains the resolved scales and region $\mathbb{S}$ contains the subgrid scales.
$k_{xC}$ and $k_{yC}$ are the cutoff wavenumbers.
$k_{xDNS}$ and $k_{yDNS}$ are the maximum wavenumbers resolved by DNS.}
\label{fig_forward_inverse_sketch}
\end{figure}

\noindent One issue of subgrid scale energy transfer modelling with eddy viscosity is that it assumes the subgrid scales only dissipate energy and neglects the energy transferred from the subgrid scales to the resolved scales, also known as `backscatter' \citep{piomelli1991subgrid}. 
Figure \ref{fig_forward_inverse_sketch} illustrates the two-way energy transfer between the resolved scales and subgrid scales. 
We aim to quantify the forward cascade and inverse cascade using mode-to-mode nonlinear energy transfer $\hat{M}$.

Recall that $\hat{M}_{(s_x,s_y)(k_x,k_y)}$ represents energy transfer from mode $(s_x,s_y)$ to mode $(k_x,k_y)$.
For the forward cascade, energy is transferred from the resolved scales in region $\mathbb{R}$ to the subgrid scales in region $\mathbb{S}$, as indicated by the solid arrow in figure \ref{fig_forward_inverse_sketch}. 
Similar to equations \eqref{eqa_N_positive} and \eqref{eqa_N_negative}, for this forward cascade, we can calculate the energy lost by each mode \eqref{eqa_N_negative_forward} in the resolved-scale region $\mathbb{R}$ and the energy gained by each mode \eqref{eqa_N_positive_forward} in the subgrid-scale region $\mathbb{S}$:
\begin{subequations}
\begin{align}
\hat{N}^{+}_{F}(k_x,k_y,n_C) &= \iint \hat{m}_{(s_x,s_y)(k_x,k_y)} \{ \hat{m}>0, (s_x,s_y) \in \mathbb{R}, (k_x,k_y) \in \mathbb{S} \}  \; \ddd s_x \ddd s_y\label{eqa_N_positive_forward} \\ 
\hat{N}^{-}_{F}(k_x,k_y,n_C) &= \iint \hat{m}_{(s_x,s_y)(k_x,k_y)} \{ \hat{m}<0, (s_x,s_y) \in \mathbb{S}, (k_x,k_y) \in \mathbb{R} \}  \; \ddd s_x \ddd s_y\label{eqa_N_negative_forward}%
\end{align}
\end{subequations}

\begin{figure}
\centering
\includegraphics[scale=0.40]{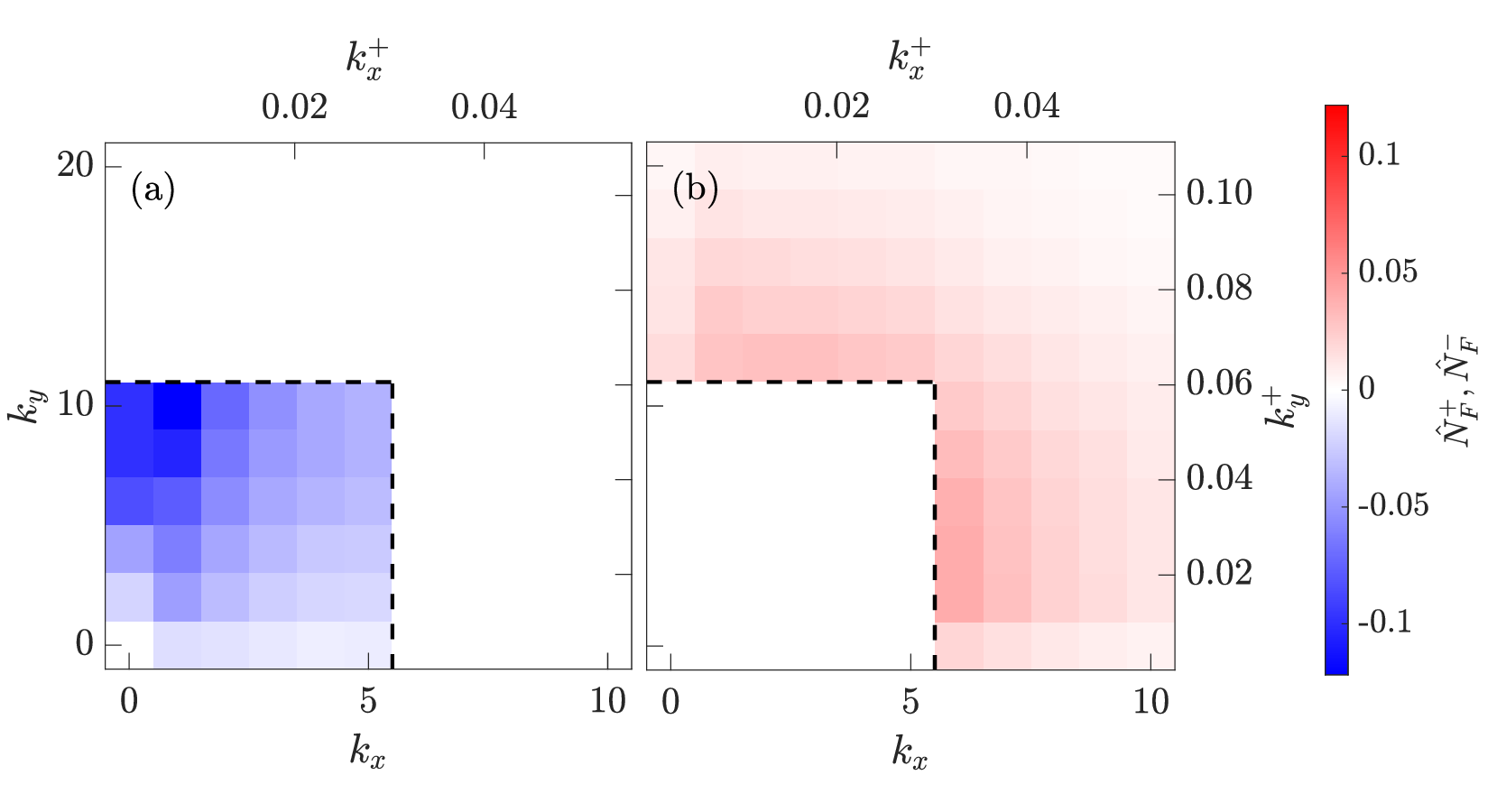}
\caption[An example of the forward cascade at $Re_{\tau}=180$.]
{An example of the forward cascade, where $n_C=5$ corresponding to $(\lambda_x^+ = 226, \lambda_y^+= 113)$ at $Re_{\tau}=180$.
(a) $\hat{N}^{-}_{F}(k_x,k_y,5)$ shows the energy lost by resolved scales in region $\mathbb{R}$; 
(b) $\hat{N}^{+}_{F}(k_x,k_y,5)$ shows the energy gained by subgrid scales in region $\mathbb{S}$.
Dashed lines mark the boundary between the resolved-scale region $\mathbb{R}$ and the subgrid-scale region $\mathbb{S}$. 
Note that the maximum wavenumbers in this figure are not the maximum wavenumbers resolved in the DNS.}
\label{fig_N_forward_example}
\end{figure}

Figure \ref{fig_N_forward_example} shows an example of the forward cascade from the resolved scales to the subgrid scales at $Re_{\tau}=180$. 
The choice of cutoff wavenumber $n_C=5$ corresponds to the filtering width $\Delta_i$ satisfying $\Delta_i \approx 5 \Delta x_i \; (i=1,2)$ at $Re_{\tau}=180$. 
Figure \ref{fig_N_forward_example}(a) shows that streamwise elongated modes with large $k_y/k_x$ in region $\mathbb{R}$ lose the most energy. 
Figure \ref{fig_N_forward_example}(b) shows that modes gaining energy in region $\mathbb{S}$ do not show significant preference with respect to $k_y/k_x$.

For the inverse cascade, energy is transferred from the subgrid scales in region $\mathbb{S}$ to the resolved scales in region $\mathbb{R}$, as indicated by the dashed-line arrow in figure \ref{fig_forward_inverse_sketch}. 
Similar to equations \eqref{eqa_N_positive} and \eqref{eqa_N_negative}, for this inverse cascade, we can calculate the energy gained by each mode \eqref{eqa_N_positive_inverse} in the resolved-scale region $\mathbb{R}$ and the energy lost by each mode \eqref{eqa_N_negative_inverse} in the subgrid-scale region $\mathbb{S}$:
\begin{subequations}
\begin{align}
\hat{N}^{+}_{I}(k_x,k_y,n_C) &= \iint \hat{m}_{(s_x,s_y)(k_x,k_y)} \{ \hat{m}>0, (s_x,s_y) \in \mathbb{S}, (k_x,k_y) \in \mathbb{R} \}  \; \ddd s_x \ddd s_y\label{eqa_N_positive_inverse} \\ 
\hat{N}^{-}_{I}(k_x,k_y,n_C) &= \iint \hat{m}_{(s_x,s_y)(k_x,k_y)} \{ \hat{m}<0, (s_x,s_y) \in \mathbb{R}, (k_x,k_y) \in \mathbb{S} \}  \; \ddd s_x \ddd s_y\label{eqa_N_negative_inverse}%
\end{align}
\end{subequations}

\begin{figure}
\centering
\includegraphics[scale=0.40]{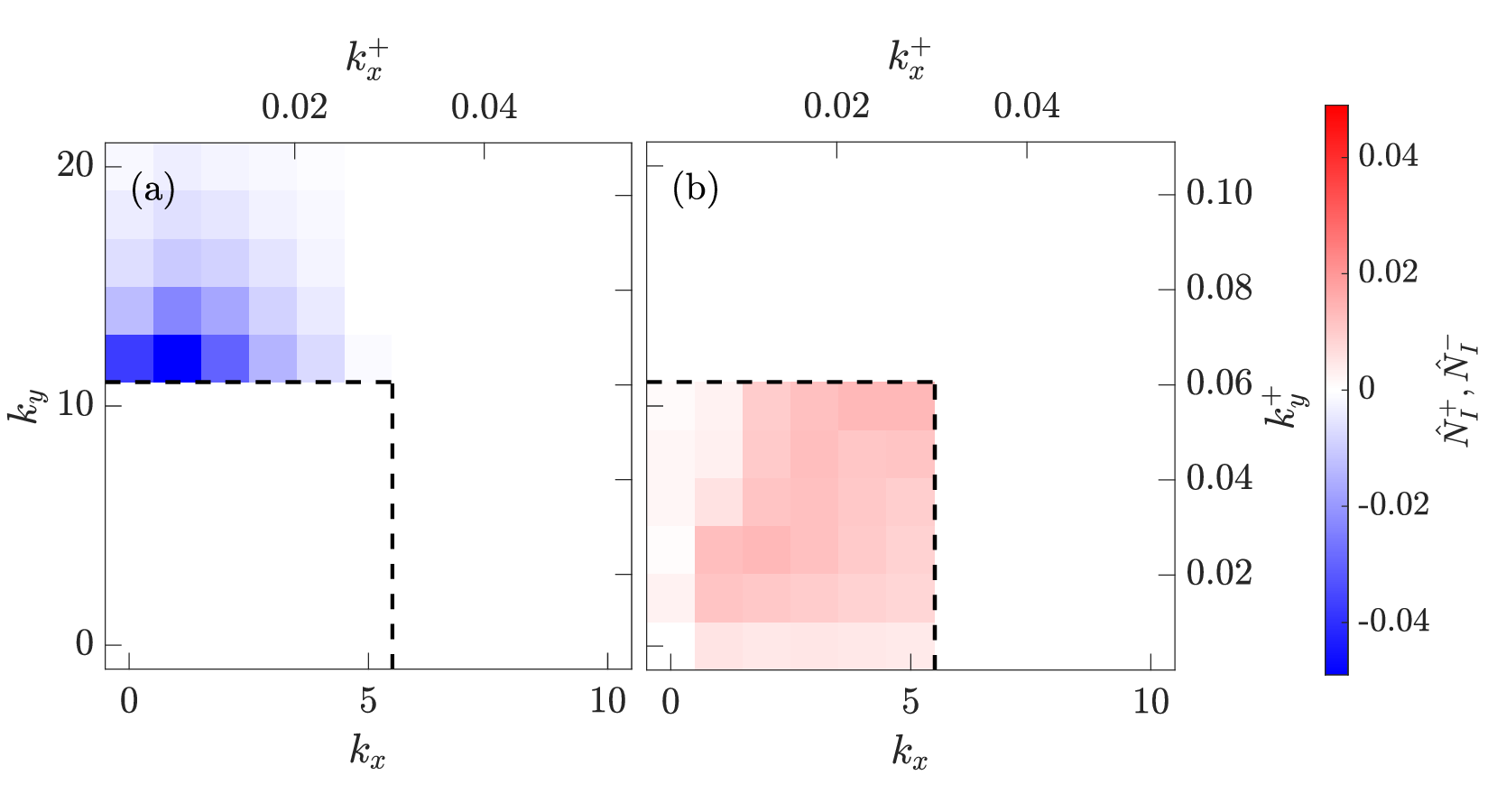}
\caption[An example of inverse cascade at $Re_{\tau}=180$.]
{Inverse cascade example: $n_C=5 (\lambda_x^+ = 226, \lambda_y^+= 113)$ at $Re_{\tau}=180$.
(a) $\hat{N}^{-}_{I}(k_x,k_y,5)$ shows how subgrid scales in region $\mathbb{S}$ lose energy; 
(b) $\hat{N}^{+}_{I}(k_x,k_y,5)$ shows how resolved scales in region $\mathbb{R}$ gain energy.}
\label{fig_N_inverse_example}
\end{figure}

Figure \ref{fig_N_inverse_example} shows an example of the inverse cascade from the subgrid scales to the resolved scales at $Re_{\tau}=180$. 
Figure \ref{fig_N_inverse_example}(a) shows the asymmetry of modes losing energy in region $\mathbb{S}$: only streamwise elongated modes with large $k_y/k_x$ lose energy.
Figure \ref{fig_N_inverse_example}(b) shows that modes gaining energy in region $\mathbb{R}$ do not show significant preference with respect to $k_y/k_x$.

We further quantify the forward cascade and inverse cascade between the resolved scales and subgrid scales determined by $n_C$:
\begin{equation}
N_{F}(n_C) = \iint_{\mathbb{S}} \hat{n}^{+}_{F} \; \ddd k_x \ddd k_y \overset{\eqref{eqa_energy_transfer_mechanism_opposite}}{=} -\iint_{\mathbb{R}} \hat{n}^{-}_{F} \; \ddd k_x \ddd k_y
    \label{eqa_NFC}
\end{equation}
\begin{equation}
N_{I}(n_C) = \iint_{\mathbb{R}} \hat{n}^{+}_{I} \; \ddd k_x \ddd k_y \overset{\eqref{eqa_energy_transfer_mechanism_opposite}}{=} -\iint_{\mathbb{S}} \hat{n}^{-}_{I} \; \ddd k_x \ddd k_y
    \label{eqa_NIC}
\end{equation}

\begin{figure}
\centering
\includegraphics[scale=0.35]{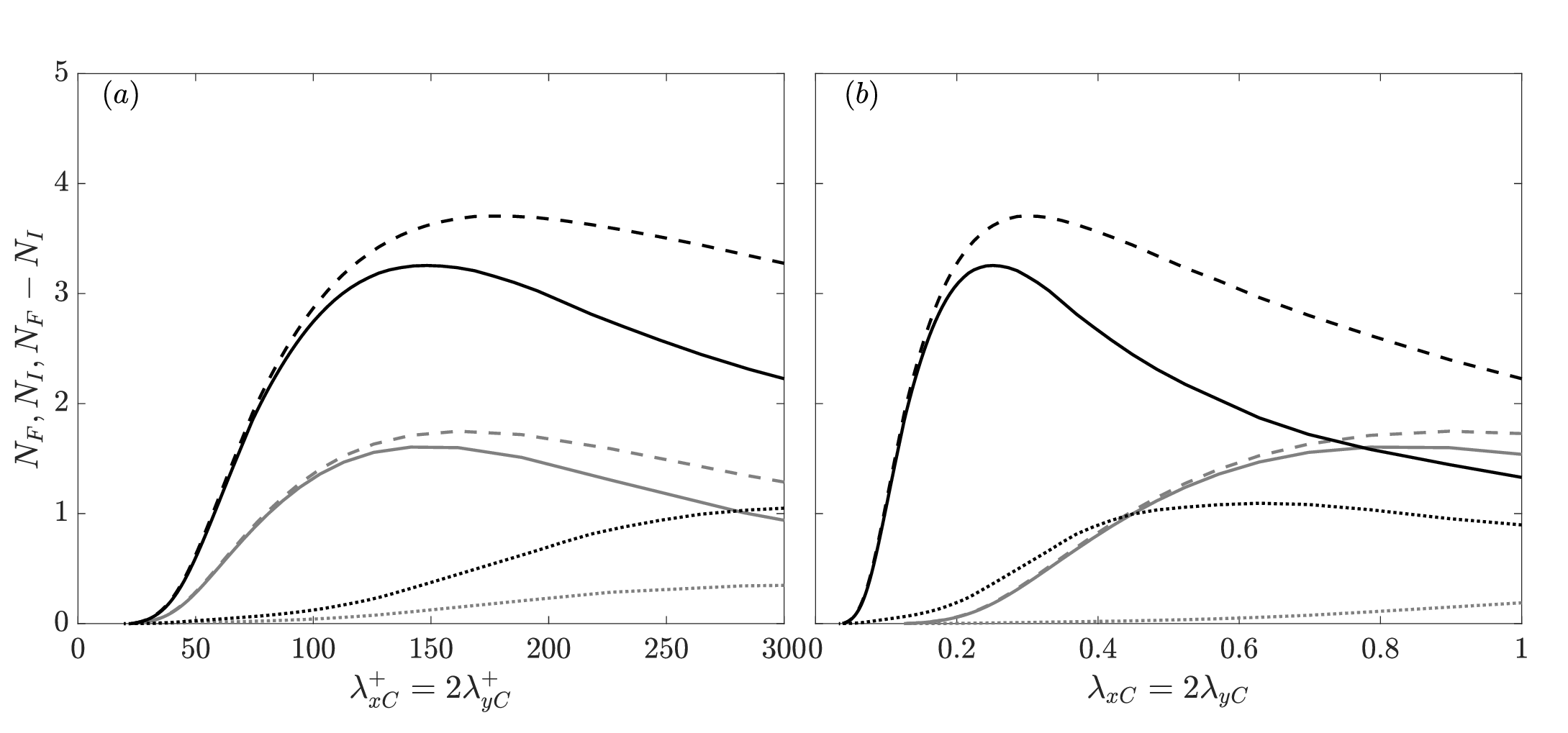}
\caption[Quantification of the forward cascade and inverse cascade at $Re_{\tau}=180$ and $590$.]
{(a) Quantification of the energy transfer between the resolved scales and subgrid scales in inner units; (b) in outer units. 
Dashed lines, forward cascade $N_{F}$; 
dotted lines, inverse cascade $N_{I}$;
solid lines, net energy cascade $N_{F}-N_{I}$. 
Grey colour, $Re_{\tau}=180$; black colour, $Re_{\tau}=590$.}
\label{fig_N_FC_IC180590_2}
\end{figure}

The forward cascade and inverse cascade calculated using equations \eqref{eqa_NFC} and \eqref{eqa_NIC} for $Re_{\tau}=180$ and $590$ are shown in figures \ref{fig_N_FC_IC180590_2}. 
A general criterion for LES is to resolve a sufficient amount of large scales which should contain 80\% of the total kinetic energy \citep{pope2000turbulent}.
By checking the DNS datasets, we know that scales with $\lambda_{x}^+ = 2 \lambda_{y}^+ >162$ need to be resolved at $Re_{\tau}=180$ and scales with $\lambda_{x}^+ = 2 \lambda_{y}^+ >232$ need to be resolved at $Re_{\tau}=590$. 
Therefore, we need to consider the energy transfer when the cut-off wavelengths satisfying $\lambda_{xC}^+ = 2 \lambda_{yC}^+ <162$ at $Re_{\tau}=180$ and $\lambda_{xC}^+ = 2 \lambda_{yC}^+ <232$ at $Re_{\tau}=590$.
We see that the forward cascade is at least four times larger than the inverse cascade when $\lambda_{xC}^+<250$, indicating that the net energy transfer is from the resolved scales to the subgrid scales and shown by the solid lines. 
This justifies why eddy viscosity considering only the forward cascade is used in LES \citep{pope2000turbulent}. 
However, the inverse cascade is not negligible when $\lambda_{xC}^+>100$, so the negligence of it has been proposed as a source of inaccuracy in LES \citep{anderson2012subgrid, cimarelli2014physics}.  

When interpreting figure \ref{fig_N_FC_IC180590_2}, two things should be noted. 
First, the results would be different if the aspect ratio between the streamwise and spanwise cutoff wavenumbers $k_{xC} / k_{yC} = \frac{1}{2}$ were changed.
Second, the above results only represent the wall-normal integrated forward and inverse cascades. 
Due to the inhomogeneity in the wall-normal direction of wall-bounded flows, the forward cascade and inverse cascade would vary with wall-normal height. 

Now, we compare the energy cascade calculated using $\hat{M}$ with the eddy viscosity method. 
The forward cascade from the resolved scales to the subgrid scales $N_{\nu}$ can be calculated using the linear eddy viscosity model \citep{pope2000turbulent}: 
\begin{equation}
N_{\nu} =  \nu \Tilde{S}^2
\label{eqa_forward_cascade_eddy_viscosity}
\end{equation}
where $\nu$ is the eddy viscosity. 
$\Tilde{S}$ is the characteristic filtered rate of strain: $\Tilde{S} = (2 \Tilde{S}_{ij} \Tilde{S}_{ij})^{\frac{1}{2}}$. 
$\Tilde{S}_{ij}$ is the filtered rate-of-strain tensor: $\Tilde{S}_{ij} = \frac{1}{2} (\frac{\partial \Tilde{u}_i}{\partial x_j} + \frac{\partial \Tilde{u}_j}{\partial x_i})$. 
$\Tilde{u}_i$ is the filtered velocity of the resolved scales. 

We can use the Smagorinsky eddy viscosity \citep{smagorinsky1963general}:
\begin{equation}
    \nu = (C_s D_z \Delta)^2 \Tilde{S}
    \label{eqa_eddy_viscosity}
\end{equation}
$C_s$ is the Smagorinsky coefficient and is set to be $C_s=0.067$ \citep{moin1982numerical}. 
$D_z$ is the damping function which accounts for low-Reynolds-number subgrid scale stress (SGS) near the wall and is set to be $D_z = 1-\mathrm{exp}(-z^+/25)$ \citep{moin1982numerical}.
Since we do not apply filtering in the wall-normal direction, the characteristic length scale of the largest subgrid scale eddies $\Delta$ is set to be $\Delta = (\Delta_1 \Delta_2)^{\frac{1}{2}}$.
After we have obtained the filtered velocity fields $\Tilde{u}_i$ using the cutoff wavenumbers \eqref{eqa_nC} in Fourier space, we calculate the forward cascade $N_{\nu}$ \eqref{eqa_forward_cascade_eddy_viscosity} and the Smagorinsky eddy viscosity $\nu$ \eqref{eqa_eddy_viscosity} directly in physical space. 

\begin{figure}
\centering
\includegraphics[scale=0.35]{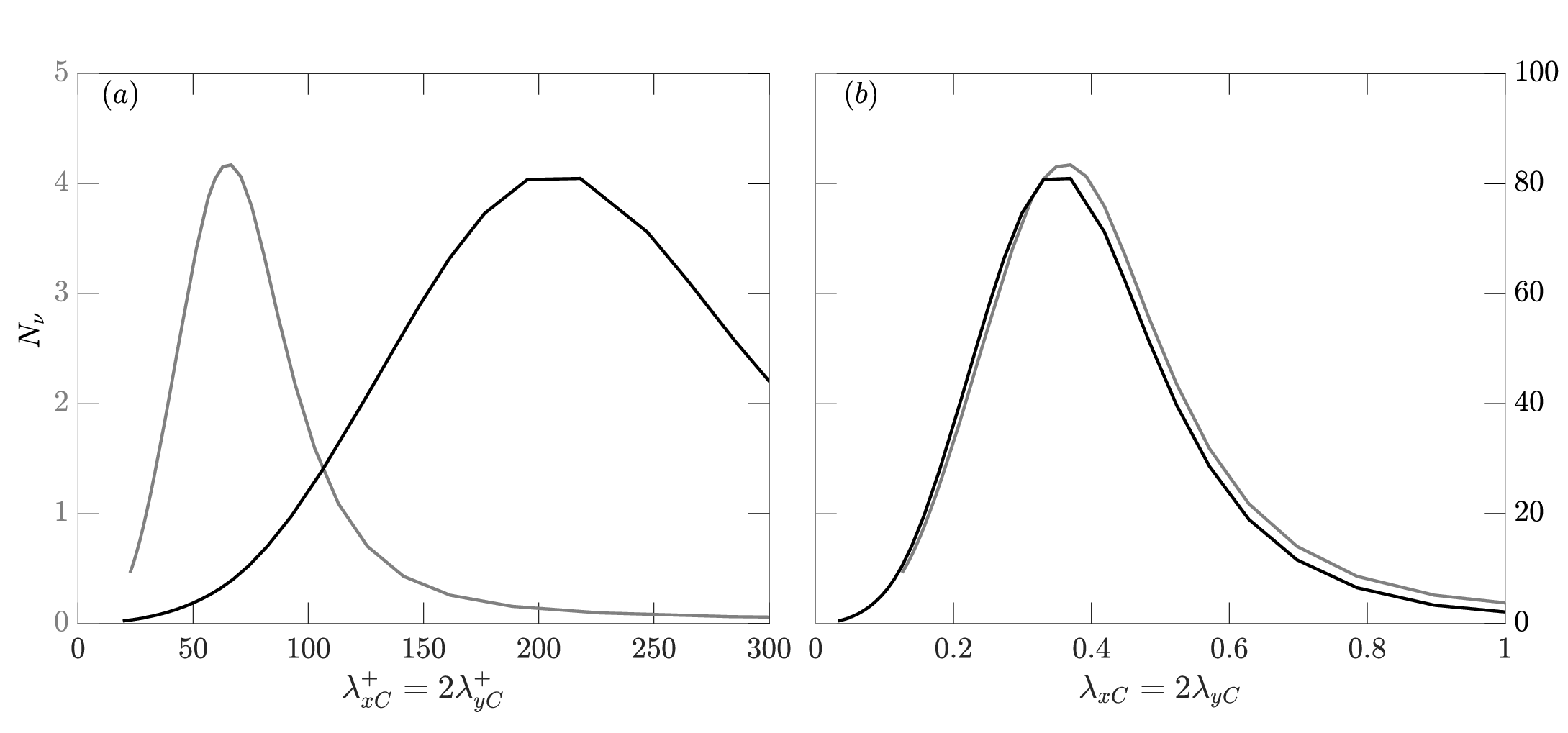}
\caption[Quantification of the forward cascade and inverse cascade at $Re_{\tau}=180$ and $590$.]
{(a) Forward cascade $N_{\nu}$ predicted by the eddy viscosity \eqref{eqa_eddy_viscosity} in inner units; (b) in outer units. 
$Re_{\tau}=180$: grey colour, left axis;
$Re_{\tau}=590$: black colour, right axis.}
\label{fig_N_forward_eddy_viscosity}
\end{figure}

Figure \ref{fig_N_forward_eddy_viscosity} shows the forward cascades from the resolved scales to subgrid scales predicted by the eddy viscosity \eqref{eqa_eddy_viscosity} at $Re_{\tau}=180$ and $590$.
We see that the forward cascade at $Re_{\tau}=590$ is approximately 20 times higher than the forward cascade at $Re_{\tau}=180$. 
The forward cascade predicted by the eddy viscosity \eqref{eqa_eddy_viscosity} first increases as the cutoff wavelengths ($\lambda_{xC}$ and $\lambda_{yC}$) decrease and then decreases to zero as the cutoff wavelengths approach the DNS grid sizes, aligning with the trends in figure \ref{fig_N_FC_IC180590_2}.
However, the values of the forward cascade predicted by the eddy viscosity \eqref{eqa_eddy_viscosity} do not match that calculated using $\hat{M}$ (the solid lines in figure \ref{fig_N_FC_IC180590_2}). 
Since $\hat{M}$ represents the wall-normal integrated energy transfer and is a time-averaged variable, statistics of energy transfer has been averaged out in the wall-normal direction and the time domain. 
That is the main reason why $N_F-N_I$ in figure \ref{fig_N_FC_IC180590_2} is substantially smaller than $N_{\nu}$ in figure \ref{fig_N_forward_eddy_viscosity}.
Nevertheless, $\hat{M}$ could be extended to include the wall-normal coordinate and $N_F-N_I$ can be calculated at each wall-normal height before being time and wall-normal averaged. 

\section{Conclusions and outlook} \label{section_conclusions}
Nonlinear energy transfer in turbulent channel flow has been investigated in Fourier space at $Re_{\tau}=180$ and $590$.
We introduce a four-dimensional variable $\hat{M}_{(s_x,s_y)(k_x,k_y)}$ which describes the nonlinear energy transfer between any two modes in streamwise-spanwise wavenumber space, by analysing the individual triadic interactions of the nonlinear energy transfer term in the spectral turbulent kinetic energy (sTKE) equation. 

We use this variable to explore three things.
First, we decompose the net nonlinear energy transfer $\hat{N}$ into its positive and negative contributions: $\hat{N}^+$ and $\hat{N}^-$.
This allows us to separately quantify the total energy gained and total energy lost by each mode. 
Second, we investigate the nonlinear energy transfer of streamwise streaks, oblique waves and TS waves. 
We observe that there exists energy transfer from streamwise-elongated structures to spanwise-elongated structures and that this transverse cascade is characterised by the aspect ratio $k_y/k_x$. 
Third, we quantify the forward cascade and inverse cascade between the resolved scales and subgrid scales in the spirit of LES.
For both Reynolds numbers considered, the forward cascade is significantly larger than the inverse cascade when $\lambda_{xC}^+<250$, justifying why eddy viscosity only considers the forward cascade is used in LES. 
However, the inverse cascade is not negligible when $\lambda_{xC}^+>100$. 
We also compare the energy cascade calculated using $\hat{M}$ with the Smagorinsky eddy viscosity.

The pathways illustrated by $\hat{M}$ represent a nonlinear energy transfer network in two-dimensional Fourier space \citep{gurcan2020turbulence}. 
A promising area for future work would be to extend mode-to-mode nonlinear energy transfer $\hat{M}_{(s_x,s_y)(k_x,k_y)}$ to include another dimension which is wall-normal coordinate to investigate the details of mode-to-mode nonlinear energy transfer at any wall-normal height.


\backsection[Acknowledgements]{This research was supported by The University of Melbourne’s Research Computing Services and the Petascale Campus Initiative.}


\backsection[Declaration of interests]{The authors report no conflict of interest.}




\appendix

\section{Full expression for $\hat{M}_{(s_x,s_y)(k_x,k_y)}$}
\label{appendix_B}
Mode-to-mode nonlinear energy transfer $\hat{M}$ is formulated from $\hat{N}$:
\begin{equation*}
    \hat{N}(k_x , k_y) =  - \langle \hat{u}_i^{(k_x , k_y)}, \widehat{\frac{\partial u_i u_j }{\partial x_j}}^{(k_x , k_y)} \rangle 
    \overset{\mathrm{continuity}}{=} 
    - \langle \hat{u}_i^{(k_x , k_y)}, \widehat{u_j \frac{\partial u_i}{\partial x_j}}^{(k_x , k_y)} \rangle
\end{equation*}
If we assign $f_i = -u_j \frac{\partial u_i}{\partial x_j}$, $\hat{N}$ could be expressed according to three different cases:
\begin{subequations}
\begin{align}
\label{eqa_N_kx_ky_1}
\begin{split}
    \underset{(k_x,k_y \in \mathbb{R^+})}{\hat{N}(k_x,k_y)} &=  (\hat{f}_i^{(k_x,k_y)}\hat{u}_i^{(-k_x,-k_y)} + \hat{f}_i^{(-k_x,k_y)}\hat{u}_i^{(k_x,-k_y)} + \mathrm{c.c.})  \\
    &=  2\mathrm{Real}\{\hat{f}_i^{(k_x,k_y)}\hat{u}_i^{(-k_x,-k_y)} + \hat{f}_i^{(-k_x,k_y)}\hat{u}_i^{(k_x,-k_y)} \}  
\end{split}
\end{align}
\begin{equation}
\label{eqa_N_0_ky_1}
    \underset{(k_y \in \mathbb{R^+})}{\hat{N}(0,k_y)} =  (\hat{f}_i^{(0,k_y)}\hat{u}_i^{(0,-k_y)} + \mathrm{c.c.}) =  2\mathrm{Real}\{\hat{f}_i^{(0,k_y)}\hat{u}_i^{(0,-k_y)} \} 
\end{equation}
\begin{equation}
\label{eqa_N_kx_0_1}
    \underset{(k_x \in \mathbb{R^+})}{\hat{N}(k_x,0)} =  (\hat{f}_i^{(k_x,0)}\hat{u}_i^{(-k_x,0)} + \mathrm{c.c.}) =  2\mathrm{Real}\{\hat{f}_i^{(k_x,0)}\hat{u}_i^{(-k_x,0)} \}
\end{equation}    
\end{subequations}
where $\mathrm{c.c.}$ represents conjugate terms. 

From the dyadic interactions in the wavenumber space, $\hat{f}_i^{(k_x,k_y)}$ could be expressed as:
\begin{align}
\begin{split}
    \hat{f}_i^{(k_x,k_y)} = &-\sum\limits_{s_x,s_y \in \mathbb{R^+}} (
    \hat{u}_j^{(k_x-s_x,k_y-s_y)}
    \widehat{\frac{\partial u_i}{\partial x_j}}^{(s_x,s_y)}+
    \hat{u}_j^{(k_x+s_x,k_y+s_y)}
    \widehat{\frac{\partial u_i}{\partial x_j}}^{(-s_x,-s_y)}\\
    &+\hat{u}_j^{(k_x+s_x,k_y-s_y)}
    \widehat{\frac{\partial u_i}{\partial x_j}}^{(-s_x,s_y)}+
    \hat{u}_j^{(k_x-s_x,k_y+s_y)}
    \widehat{\frac{\partial u_i}{\partial x_j}}^{(s_x,-s_y)})\\
    &-\sum\limits_{b \in \mathbb{R^+}} (
    \hat{u}_j^{(k_x,k_y-s_y)}
    \widehat{\frac{\partial u_i}{\partial x_j}}^{(0,s_y)}+
    \hat{u}_j^{(k_x,k_y+s_y)}
    \widehat{\frac{\partial u_i}{\partial x_j}}^{(0,-s_y)})\\
    &- \sum\limits_{a \in \mathbb{R^+}} (
    \hat{u}_j^{(k_x-s_x,k_y)}
    \widehat{\frac{\partial u_i}{\partial x_j}}^{(s_x,0)}+
    \hat{u}_j^{(k_x+s_x,k_y)}
    \widehat{\frac{\partial u_i}{\partial x_j}}^{(-s_x,0)})
\end{split}
\label{eqa_fi}
\end{align}

Substitute equation \eqref{eqa_fi} into equations \eqref{eqa_N_kx_ky_1}, \eqref{eqa_N_0_ky_1} and \eqref{eqa_N_kx_0_1}.
The mode-to-mode nonlinear transfer between two modes for nine different cases is summarised here:
\begin{subequations}
\begin{align}
\label{eqa_a_b_kx_ky}
\begin{split}
    &\hat{M}_{(s_x,s_y)(k_x,k_y)} =   -2\mathrm{Real}\{ \\  
    & \hat{u}_j^{(k_x-s_x,k_y-s_y)}
    \widehat{\frac{\partial u_i}{\partial x_j}}^{(s_x,s_y)}
    \hat{u}_i^{(-k_x,-k_y)}+
    \hat{u}_j^{(k_x+s_x,k_y+s_y)}
    \widehat{\frac{\partial u_i}{\partial x_j}}^{(-s_x,-s_y)}
    \hat{u}_i^{(-k_x,-k_y)}+\\
    &\hat{u}_j^{(k_x+s_x,k_y-s_y)}
    \widehat{\frac{\partial u_i}{\partial x_j}}^{(-s_x,s_y)}
    \hat{u}_i^{(-k_x,-k_y)}+
    \hat{u}_j^{(k_x-s_x,k_y+s_y)}
    \widehat{\frac{\partial u_i}{\partial x_j}}^{(s_x,-s_y)}
    \hat{u}_i^{(-k_x,-k_y)}+\\
    &\hat{u}_j^{(-k_x-s_x,k_y-s_y)}
    \widehat{\frac{\partial u_i}{\partial x_j}}^{(s_x,s_y)}
    \hat{u}_i^{(k_x,-k_y)}+
    \hat{u}_j^{(-k_x+s_x,k_y+s_y)}
    \widehat{\frac{\partial u_i}{\partial x_j}}^{(-s_x,-s_y)}
    \hat{u}_i^{(k_x,-k_y)}+\\
    &\hat{u}_j^{(-k_x+s_x,k_y-s_y)}
    \widehat{\frac{\partial u_i}{\partial x_j}}^{(-s_x,s_y)}
    \hat{u}_i^{(k_x,-k_y)}+
    \hat{u}_j^{(-k_x-s_x,k_y+s_y)}
    \widehat{\frac{\partial u_i}{\partial x_j}}^{(s_x,-s_y)}
    \hat{u}_i^{(k_x,-k_y)}\} 
\end{split}
\end{align}

\begin{align}
\label{eqa_0_b_kx_ky}
\begin{split}
    &\hat{M}_{(0,s_y)(k_x,k_y)} = -2\mathrm{Real} \{ \\
    &\hat{u}_j^{(k_x,k_y-s_y)}
    \widehat{\frac{\partial u_i}{\partial x_j}}^{(0,s_y)}
    \hat{u}_i^{(-k_x,-k_y)}+
    \hat{u}_j^{(k_x,k_y+s_y)}
    \widehat{\frac{\partial u_i}{\partial x_j}}^{(0,-s_y)}
    \hat{u}_i^{(-k_x,-k_y)} +\\
    &\hat{u}_j^{(-k_x,k_y-s_y)}
    \widehat{\frac{\partial u_i}{\partial x_j}}^{(0,s_y)}
    \hat{u}_i^{(k_x,-k_y)}+
    \hat{u}_j^{(-k_x,k_y+s_y)}
    \widehat{\frac{\partial u_i}{\partial x_j}}^{(0,-s_y)}
    \hat{u}_i^{(k_x,-k_y)} \} 
\end{split}
\end{align}

\begin{align}
\label{eqa_a_0_kx_ky}
\begin{split}
    &\hat{M}_{(s_x,0)(k_x,k_y)} =  -2\mathrm{Real} \{ \\
    &\hat{u}_j^{(k_x-s_x,k_y)}
    \widehat{\frac{\partial u_i}{\partial x_j}}^{(s_x,0)}
    \hat{u}_i^{(-k_x,-k_y)}+
    \hat{u}_j^{(k_x+s_x,k_y)}
    \widehat{\frac{\partial u_i}{\partial x_j}}^{(-s_x,0)}
    \hat{u}_i^{(-k_x,-k_y)}+\\
    &\hat{u}_j^{(-k_x-s_x,k_y)}
    \widehat{\frac{\partial u_i}{\partial x_j}}^{(s_x,0)}
    \hat{u}_i^{(k_x,-k_y)}+
    \hat{u}_j^{(-k_x+s_x,k_y)}
    \widehat{\frac{\partial u_i}{\partial x_j}}^{(-s_x,0)}
    \hat{u}_i^{(k_x,-k_y)} \} 
\end{split}
\end{align}

\begin{align}
\label{eqa_a_b_0_ky}
\begin{split}
    &\hat{M}_{(s_x,s_y)(0,k_y)} =  -2\mathrm{Real}\{ \\
    &\hat{u}_j^{(-s_x,k_y-s_y)}
    \widehat{\frac{\partial u_i}{\partial x_j}}^{(s_x,s_y)}
    \hat{u}_i^{(0,-k_y)}+
    \hat{u}_j^{(s_x,k_y+s_y)}
    \widehat{\frac{\partial u_i}{\partial x_j}}^{(-s_x,-s_y)}
    \hat{u}_i^{(0,-k_y)}\\
    &+\hat{u}_j^{(s_x,k_y-s_y)}
    \widehat{\frac{\partial u_i}{\partial x_j}}^{(-s_x,s_y)}
    \hat{u}_i^{(0,-k_y)}+
    \hat{u}_j^{(-s_x,k_y+s_y)}
    \widehat{\frac{\partial u_i}{\partial x_j}}^{(s_x,-s_y)}
    \hat{u}_i^{(0,-k_y)} \} 
\end{split}
\end{align}

\begin{align}
\label{eqa_0_b_0_ky}
\begin{split}
    &\hat{M}_{(0,s_y)(0,k_y)} =  -2\mathrm{Real}\{ 
    \hat{u}_j^{(0,k_y-s_y)}
    \widehat{\frac{\partial u_i}{\partial x_j}}^{(0,s_y)}
    \hat{u}_i^{(0,-k_y)}+
    \hat{u}_j^{(0,k_y+s_y)}
    \widehat{\frac{\partial u_i}{\partial x_j}}^{(0,-s_y)}
    \hat{u}_i^{(0,-k_y)} \} 
\end{split}
\end{align}

\begin{align}
\label{eqa_a_0_0_ky}
\begin{split}
    &\hat{M}_{(s_x,0)(0,k_y)} =  -2\mathrm{Real}\{ 
    \hat{u}_j^{(-s_x,k_y)}
    \widehat{\frac{\partial u_i}{\partial x_j}}^{(s_x,0)}
    \hat{u}_i^{(0,-k_y)}+
    \hat{u}_j^{(s_x,k_y)}
    \widehat{\frac{\partial u_i}{\partial x_j}}^{(-s_x,0)}
    \hat{u}_i^{(0,-k_y)}\} 
\end{split}
\end{align}

\begin{align}
\label{eqa_a_b_kx_0}
\begin{split}
    &\hat{M}_{(s_x,s_y)(k_x,0)} = -2\mathrm{Real}\{ \\
    &\hat{u}_j^{(k_x-s_x,-s_y)}
    \widehat{\frac{\partial u_i}{\partial x_j}}^{(s_x,s_y)}
    \hat{u}_i^{(-k_x,0)}+
    \hat{u}_j^{(k_x+s_x,s_y)}
    \widehat{\frac{\partial u_i}{\partial x_j}}^{(-s_x,-s_y)}
    \hat{u}_i^{(-k_x,0)}\\
    &+\hat{u}_j^{(k_x+s_x,-s_y)}
    \widehat{\frac{\partial u_i}{\partial x_j}}^{(-s_x,s_y)}
    \hat{u}_i^{(-k_x,0)}+
    \hat{u}_j^{(k_x-s_x,s_y)}
    \widehat{\frac{\partial u_i}{\partial x_j}}^{(s_x,-s_y)}
    \hat{u}_i^{(-k_x,0)} \} 
\end{split}
\end{align}

\begin{align}
\label{eqa_0_b_kx_0}
\begin{split}
    &\hat{M}_{(0,s_y)(k_x,0)} = -2\mathrm{Real}\{ 
    \hat{u}_j^{(k_x,-s_y)}
    \widehat{\frac{\partial u_i}{\partial x_j}}^{(0,s_y)}
    \hat{u}_i^{(-k_x,0)}+
    \hat{u}_j^{(k_x,s_y)}
    \widehat{\frac{\partial u_i}{\partial x_j}}^{(0,-s_y)}
    \hat{u}_i^{(-k_x,0)} \}
\end{split}
\end{align}

\begin{align}
\label{eqa_a_0_kx_0}
\begin{split}
    &\hat{M}_{(s_x,s_y)(k_x,0)} = -2\mathrm{Real}\{ 
    \hat{u}_j^{(k_x-s_x,0)}
    \widehat{\frac{\partial u_i}{\partial x_j}}^{(s_x,0)}
    \hat{u}_i^{(-k_x,0)}+
    \hat{u}_j^{(k_x+s_x,0)}
    \widehat{\frac{\partial u_i}{\partial x_j}}^{(-s_x,0)}
    \hat{u}_i^{(-k_x,0)}\}
\end{split}
\end{align}
\end{subequations}

\bibliographystyle{jfm}
\bibliography{jfm}

\end{document}